\documentclass[twocolumn]{aastex63}
\usepackage{lipsum} 

\usepackage{graphicx}
\usepackage{tikz}
\usepackage[export]{}

\usepackage{amsmath}

\usepackage{fontenc}
\usepackage{cancel}
\usepackage{tensor}
\usepackage{rotating} 

\renewcommand\vec{\mathbf}
\DeclareMathSymbol{\mh}{\mathord}{operators}{`\-}
\received{March -, 2024}
\revised{-- --, 2024}
\accepted{01-Jul, 2024} %{\today}
\submitjournal{APJ}

\shorttitle{Turbulent-like fluctuations and the formation of plasmoids}
\shortauthors{J.A. Agudelo Rueda et al.}

\graphicspath{{./}{figures/}}
\begin{document}

\title{On the Effect of Driving Turbulent-like Fluctuations on a Harris-Current Sheet Configuration and the Formation of Plasmoids.}

\correspondingauthor{Jeffersson Andres Agudelo Rueda}
\email{jeffersson.agudelo@northumbria.ac.uk}

\author[0000-0001-5045-0323]{Jeffersson A Agudelo Rueda}
\affiliation{Department of Physics and Astronomy, Dartmouth College, Hanover, NH, USA}
\affiliation{Department: Mathematics, Physics and Electrical Engineering, Northumbria University, Newcastle Upon Tyne, UK}
%\affiliation{Mullard Space Science Laboratory, UCL,
%Holmbury St. Mary, Dorking, Surrey, RH5 6NT, UK}

\author[0000-0001-5880-2645]{Yi-Hsin Liu}
\affiliation{Department of Physics and Astronomy, Dartmouth College, Hanover, NH, USA}

\author[0000-0002-8495-6354]{Kai Germaschewski}
\affiliation{Space Science Center,
University of New Hampshire,
Durham NH 03824, USA}

\author[0000-0003-0377-9673]{Michael Hesse}
\affiliation{NASA,
Goddard Space Flight Center,
Greenbelt, Maryland, USA}

\author[0000-0002-5431-174X]{Naoki Bessho}
\affiliation{Department of Astronomy,
University of Maryland,
College Park, Maryland, USA}

%-----------------------------------------------------------------
%What is specifically the process that is affecting the tearing. The distribution function give an inside but the argument is not conclusive. Needs to be tied to the tearing physics and the reconnection physics. 

%-----------------------------------------------------------------

%\author[0000-0001-5278-8029]{Xiaocan Li}
%\affiliation{Department of Physics and Astronomy, Dartmouth College, Hanover, NH, USA}

% This abstract needs to be corrected
%-----------------------------------------------------------------
\begin{abstract}
Energy dissipation in collisionless plasmas is one of the most outstanding open questions in plasma physics. Magnetic reconnection and turbulence are two phenomena that can produce the conditions for energy dissipation. These two phenomena are closely related to each other in a wide range of plasmas. Turbulent fluctuations can emerge in critical regions of reconnection events, and magnetic reconnection can occur as a product of the turbulent cascade. In this study, we perform 2D particle-in-cell simulations of a reconnecting Harris current sheet in the presence of turbulent fluctuations to explore the effect of turbulence on the reconnection process in collisionless non-relativistic pair-plasmas. We find that the presence of a turbulent field can affect the onset and evolution of magnetic reconnection. Moreover, we observe the existence of a scale dependent amplitude of magnetic field fluctuations above which these fluctuations are able to disrupt the growing of magnetic islands. These fluctuations provide thermal energy to the particles within the current sheet and preferential perpendicular thermal energy to the background population. 
\end{abstract}
%----------------------------------------------------------------

%----------------------------------------------------------------
%------------------------------------------- ---------------------

%---------------------------------------------------------------
\section{Introduction} 
\label{sec:intro}

%magnetic reconnection
Magnetic reconnection is a process in which the magnetic-field lines reconnect changing the magnetic-field topology while converting magnetic energy into the plasma particles kinetic and thermal energy \citep{vasyliunas1975theoretical,parker1957acceleration,sweet195814,dungey1961interplanetary}. It occurs in a wide range of plasmas, e.g., solar wind \citep{gosling2012magnetic,phan2020parker}, black hole accretion disks \citep[and references therein]{ripperda2020magnetic, zweibel2016}. Likewise, most plasmas develop turbulent states in which the energy is distributed among a spectrum of spatial and temporal scales. For instance, the solar wind \citep{coleman1968turbulence, marsch1990spectral,martinovic2021multiscale}, the plasma sheet in the Earth's magnetosphere \citep{borovsky2003mhd}, the Earth's magnetotail \citep{ergun2022observation}, and active galactic nuclei (AGN) \citep{molokov2007turbulence,howes2008kinetic} are in the state of turbulence. Turbulence manifest itself as an energy transfer process across a broad range of spatial scales that involve non-linear temporal scales \citep{tsytovich1977theory,yamada2008anatomy}. Magnetic reconnection and turbulence are intimately related. For instance, in the Magnetohydrodynamic (MHD) regime, introducing turbulent fluctuations in a reconnecting current sheet enhances the reconnection rate (rate of reconnecting magnetic flux) compared to the laminar steady reconnection case both in 2D \citep{matthaeus1986turbulent,loureiro2009turbulent} and in 3D \citep{lazarian1999reconnection,kowal2009numerical}. 

Moreover, kinetic simulations \citep{adhikari2020reconnection,adhikari2024scale} and observations of magnetic reconnection events at the Earth's magnetosphere show the presence of turbulence that develops at the reconnection jets \citep{eastwood2009observations,richard2024turbulence}.
Additionally, the presence of turbulent fluctuations can provide the conditions for the tearing instability \citep{galeev1976tearing,drake1977kinetic} to grow in the magnetotail \citep{esarey1987turbulent,liu2014onset}. 

On the other hand, plasma turbulence creates reconnection sites \citep{servidio2009magnetic} and thin current sheets that are unstable to the tearing instability. The reconnection of these current sheets drives sub-ion-scales cascades \citep{franci2017magnetic,manzini2023subion} and the disruption of the current sheet modifies the cross-scale energy transfer \citep{boldyrev2017magnetohydrodynamic,boldyrev2019role,boldyrev2020tearing}.

%{Role of reconnection in inertial kinetic Alfvén wave turbulence \citep{boldyrev2019role} Magnetic reconnection in two-dimensional magnetohydrodynamic turbulence \citep{servidio2009magnetic} Observations of turbulence generated by magnetic reconnection \citep{eastwood2009observations} Magnetic reconnection as a driver for a sub-ion-scale cascade in plasma turbulence \citep{franci2017magnetic} Subion-Scale Turbulence Driven by Magnetic Reconnection \citep{manzini2023subion} Turbulence in Magnetic Reconnection Jets from Injection to Sub-Ion Scales \citep{richard2024turbulence} Scale filtering analysis of kinetic reconnection and its associated turbulence \citep{adhikari2024scale}}
 
%pair plasmas in context
Electron-positron plasmas, also known as pair plasmas, are present in astrophysical jets \citep{kundt1987astrophysical,romero2021content}, gamma-ray burst from active galactic nuclei (AGN) \citep{zdziarski1990electron,henri1991relativistic,henri1993gamma}, super novas \citep{woosley2005physics,woosley2007pulsational} and black-hole accretion disks \citep{khiali2016high}. Studying fundamental processes (e.g., magnetic reconnection and turbulence) in pair plasmas is important to understand energising mechanisms in astrophysical objects \citep{guo2015particle,pucci2020onset} and in the non-relativistic regime \citep{bessho2005collisionless,bessho2007fast,hesse2007dissipation}. Given the lack of direct in-situ measurements in pair plasmas, numerical simulations present themselves as a powerful tool to explore magnetic reconnection \citep{hesse2007dissipation,bessho2007fast,liu2015scaling} and plasma turbulence \citep{zocco2017slab,loureiro2018turbulence} in this type of plasmas. Moreover, the development of electron-positron plasma devises is an ongoing effort towards the understanding of fundamental plasma processes \citep{pedersen2012plans,sarri2015generation}.

A fundamental part of the onset of magnetic reconnection is the formation of magnetic islands, i.e., plasmoids formation in a thin current sheet by means of the tearing instability \citep{drake1977kinetic,karimabadi2005physics}. Although a broad range of fluctuations can provide free energy sources for the tearing instability to grow, it is not clear which range of scales comes into play to either trigger or modify the evolution of the plasmoids. Thus, in this study we test the effect of driving fluctuations at different spatial and temporal scales. Particularly, the questions we address in this paper are: \emph{What is the effect of driving fluctuations on the onset of the tearing modes?} And, \emph{how can the growing of magnetic islands be changed by these fluctuations?} 

We show here that, in a 2D case and for kinetic current sheets, only driven magnetic-field fluctuations with amplitudes larger than a critical amplitude $\delta B_{c}$ and with wavelengths comparable to or smaller than half the initial thickness of the current sheet, are able to suppress the growing of the of magnetic islands. We observe that $\delta B_{c}$ depends on the wavelength of the driven fluctuations. In section \ref{sec:methodology} we describe our numerical set-up and the configuration of the simulations that we perform. In section \ref{sec:simulation_results} we present the results and in section \ref{sec:discussion} we discuss our results. Finally in section \ref{sec:conclusions} we conclude and provide guidelines for future work.

%---------------------------------------------------------------------
\section{Methodology} 
\label{sec:methodology}

%----------------------------------------------------------------------
We use the Plasma Simulation Code \citep[PSC,][]{germaschewski2016plasma}, an explicit Particle-In-Cell (PIC) code, to study the effect of turbulent-like fluctuations on the magnetic reconnection dynamics of a Harris Current-Sheet \citep{harris1962plasma} in 2D. 

%---------------------------------------------------------------------

%----------------------------------------------------------------------
\subsection{Harris Current Sheet Configuration}

To set the Harris current sheet equilibrium \citep{harris1962plasma} we use the magnetic field 

\begin{align}
    \vec{B} = B_{g}\hat{x} +  B_{0} \tanh\left(\frac{y}{\Delta}\right) \hat{z}
\end{align}

\noindent and the density of the particle species $s=i,e$ 

\begin{align}
n_{s} = n_{s,0} cosh^{-2}\left(\frac{y}{\Delta}\right) + n_{b},
\end{align}

\noindent where $B_{0}$ is the asymptotic value of the upstream magnetic field, $B_{g}$ is the guide field, $\Delta$ is the initial half-thickness of the central current sheet, $n_{s,0}$ is the density of the main population at the central current sheet and $n_{b}=n_{s,b}$ is the background density of the species $s$. We take $n_{i,0} = n_{e,0} = n_{0}$ and $n_{i,b} = n_{e,b} = n_{b}$ and set 

\begin{align}
    n_{0}k_{B}(T_{e,0} + T_{i,0}) = \frac{B_{0}^{2}}{2\mu_{0}},
    \label{eqn:temp_e_b}
\end{align}

to keep the pressure balance, where {$k_{B}$ is the Boltzmann's constant}, $T_{e,0}$ and $T_{i,0}$ are the temperature of the electron and ion populations at the centre of the current sheet respectively. We set the electron drift velocity $v_{e,d} = - 2k_{B}T_{e,0}/eB_{0}\Delta$ and $v_{i,d}=-v_{e,d}T_{i,0}/T_{e,0}$ for the main populations as it is required for the standard Harris equilibrium. We define background plasma beta

\begin{align}
    \beta_{b} = \frac{2 \mu_{0} n_{b}k_{B}T_{e,b} (1 + T_{i,b}/T_{e,b})}{B_{0}^{2}}
    \label{eqn:beta}
\end{align}

where $T_{e,b}$ and $T_{i,b}$ are the temperature of the background electron and background ion populations respectively. 

%----------------------------------------------------------------------
\subsection{Driving magnetic-field fluctuations}
\label{sec:inject_fluc}

To drive magnetic fluctuations into a simulation domain of size $L_{y} \times L_{z}$, we use the Langevin-antenna \citep{tenbarge2014oscillating} method and follow \cite{grovselj2019fully} to implement it on our PIC simulations. This method has been applied to study kinetic turbulence in both classical \citep{grovselj2019kinetic} and relativistic \citep{zhdankin2017kinetic,zhdankin2018system,zhdankin2019electron} plasmas.

% Equations
%---------------------------------------------------
To self-consistently drive magnetic fluctuations $\delta \vec{B}_{\mathbf{ext}}$, we consider the magnetic vector potential in the out-of-plane direction $\delta \vec{A}=\delta A_{x} \hat{x}$ where

\begin{align}
    \delta A_{x}=Real\left[ \sum_{j=1}^{N-modes}\frac{b_{\nu,k_{j}}}{k_{j,\perp}}e^{i(\vec{k}_{j} \cdot \vec{r})}\right].
    \label{eqn:vec_Az}
\end{align}

\noindent Here the subscript $\nu$ represents the time step, $\vec{r}=(y,z)$ is the position, $\vec{k}_{j}=(k_{j,y},k_{j,z})$ is the wavevector of the $j$-$th$ fluctuation whose components are 

\begin{align}
    k_{j,y} = \frac{m_{y,j}2\pi}{L_{y}} \ ; \ k_{j,z} = \frac{m_{z,j}2\pi}{L_{z}},
\end{align}

%${k}_{m,j,\perp}=\sqrt{k_{m,j,y}^{2} + k_{m,j,z}^{2}}$
\noindent where integers $m_{y}$ and $m_{z}$ represent the mode order in each direction and ${k}_{j,\perp}$ is the wavenumber perpendicular to the local magnetic field $\vec{B}_{l} = B_{g}\hat{x} + B_{0}\tanh{(y/\Delta)}\hat{z}$. Thus, the external magnetic field is $\delta \vec{B}_{\mathbf{ext}} = \nabla \times \delta \vec{A}$. We drive the fluctuations via the external current 

\begin{equation}
\delta \vec{J}_{\mathbf{ext}} = \frac{1}{\mu_{0}}\nabla \times \delta \vec{B}_{\mathbf{ext}}. 
    \label{eqn:Jext}
\end{equation}

This current is added to the self-consistent current $\vec{J}$ and we use the total current $\vec{J}_{T}=\vec{J} + \delta \vec{J}_{\mathbf{ext}}$ to advance the electric field $\vec{E}$ at each time step. The coefficients $b_{\nu,k_{j}}$ are updated at every time step as

\begin{align}
    b_{0,k_{j}} &= \delta B_{0}e^{i\phi_{k_{j}}}, \nonumber\\
    b_{\nu+1,k_{j}} &= b_{\nu,k_{j}}e^{-(i\omega_{0}+\gamma_{0})\Delta t} + \delta B_{0}\sqrt{12\gamma_{0}\Delta t u_{\nu,k_{j}}} \nonumber,
\end{align}

\noindent where $\delta B_{0}$ is the amplitude of the fluctuations, $\phi_{k_{j}}$ is the random phase of the $j$-th fluctuation, $\omega_{0}$ is the driving frequency, $\gamma_{0}$ is the decay rate, $\Delta t$ is the time step and $u_{\nu,k_{j}}$ is a random number in the interval $[-0.5,0.5]$. We set a polarization such that the fluctuations are $\delta B_{y}=\delta B_{y}(z)$, $\delta B_{z}=\delta B_{z}(y)$, and $\delta B_{x}=0$ to enforce $\nabla \cdot \vec{B} = 0$.

%---------------------------------------------------------------------
\subsection{Numerical set up}
\label{sec:num_set}

%---------------------------------------------------
The normalization parameters for our study are the electron plasma frequency $\omega_{pe}=\sqrt{n_{e,0}e^2/m_{e}\epsilon_{0}}$ and the electron inertial length $d_{e}=c/\omega_{pe}$, where $e$ is the elementary charge, $m_{e}$ is the electron mass, $\epsilon_{0}$ is the electric permeability, $c$ is the speed of light and $n_{e,0}$ is the electron density at the centre of the current sheet. {The length is in units of $d_{e}$, the time in units of $\omega_{pe}^{-1}$, the magnetic field is normalized to the upstream magnetic field $B_{0}$, the density is normalized to $n_{e,0}$ and the particle velocities are in units of $c$}. {We simulate an electron-positron plasma, i.e., $m_{i}/m_{e}=1$ and $d_{i}=d_{e}$}, in the plane $ L_{y} \times L_{z} = 128 d_{e}\times 256 d_{e}$ with spatial resolution $\Delta y = \Delta z = 0.11d_{e}$, where $d_{i}=c/\omega_{pi}$ is the positron inertial length, $m_{i}$ is the positron mass and $\omega_{pi}=\sqrt{n_{0}e^2/\epsilon_{0}m_{i}}$ is the positron plasma frequency. We use 400 macro-particles per cell and set $\beta_{b}=0.2$, $\Delta=4d_{e}$, $T_{i,b} = T_{e,b}$, $T_{i,0} = T_{e,0}$, $n_{i,b} = n_{e,b}$, $n_{i,0} = n_{e,0}$, the ratio $n_{e,0}/n_{e,b}$ can be found in table \ref{tab:com_cases} and $\omega_{pe}/\Omega_{e} = 5$ where $\Omega_{e} = eB_{0}/m_{e}$ is the electron gyrofrequency.   
%---------------------------------------------------------------------

%and Debye length $\lambda_{D}=0.14d_{i}$

%--------------------------------------------------------------
\subsection{Description of the different runs}

In this section we describe the different runs that we test in this study. Run $SA$ corresponds to a Harris Current-Sheet configuration with no initial perturbation. Thus, the onset of the tearing instability arises spontaneously. Therefore, this is the run we use to contrast with. Run $SB$ corresponds to the same configuration as run $SA$ plus an additional initial large-wavelength perturbation

\begin{align}
    \delta B_{y} = \delta B_{y,0} \sin\left[\frac{2\pi}{L_{z}}\left(z-\frac{L_{z}}{2}\right)\right] \cos\left(\frac{\pi y}{L_{y}} \right), \\
    \delta B_{z} = \delta B_{z,0} \cos\left[\frac{2\pi}{L_{z}}\left(z-\frac{L_{z}}{2}\right)\right] \sin\left(\frac{\pi y}{L_{y}} \right), 
\end{align}
 
\noindent where $\delta B_{y,0} = - \delta B 2\pi \Delta/L_{z}$ and $\delta B_{z,0} = \delta B \pi \Delta/L_{y}$, where we use $\delta B =0.2B_{0}$. This initial perturbation, which we refer to as central pinch, is often used to speed up the system to the steady reconnection state. For all other runs, except for run $SF$, which is similar to run $SA$ but with a larger guide field $B_{g}=0.2 B_{0}$, we set a Harris Current-Sheet configuration and constantly drive magnetic fluctuations using the aforementioned \citep[Langevin-antenna,][]{tenbarge2014oscillating} method (Section \ref{sec:inject_fluc}). For run $SC$ we drive a superposition of 8 modes with the same amplitude ($ \langle |\delta B_{y}|/|\mathbf{B}| \rangle = 0.002$), with the same wavemode $m_{y,j} = m_{z,j} = 4$ and with the same frequency $\omega_{0}/\omega_{pe}=0.004$, where $|\mathbf{B}|$ is the magnitude of the local magnetic field and $\langle ... \rangle$ is the average over the entire simulation domain. 

For run $SD$ we drive a superposition of 32 modes of equal amplitude ($\langle |\delta B_{y}|/|\mathbf{B}| \rangle = 0.004$), with different wavemodes $m_{y,j} = m_{z,j} = 1,2,3,4$ and with the same frequency $\omega_{0}/\omega_{pe}=0.004$. For run $SE$ we use a larger amplitude for the fluctuations ($\langle |\delta B_{y}|/|\mathbf{B}| \rangle = 0.016$), a slightly higher background density ($n_{b}=0.2n_{0}$) a higher guide field ($B_{g}=0.2B_{0}$), wavemodes $m_{y,j} = 1,2,3,4$ and $m_{z,j} = 2,4,6,8$, and with the same frequency $\omega_{0}/\omega_{pe}=0.005$. In all runs where we drive magnetic fluctuations ($SC-SN$), the amplitude of the fluctuations is determined by 

\begin{align}
\frac{\delta B_{y}}{B_{0}}= \frac{\delta B_{z}}{B_{0}}= \frac{\delta B}{B_{0}} \frac{L_{y}}{ Lz}.
\end{align}

Note that $\delta B_{y}$ and $\delta B_{z}$ for run $SB$ are different than $\delta B_{y}$ and $\delta B_{z}$ for runs $SC-SN$. In runs $SG$ and $SH$, we drive 32 wavemodes $m_{y,j} = 1,2,3,4$ and $m_{z,j} = 2,4,6,8$ with frequency ($\omega_{0}/\omega_{pe} = 0.2$) and with amplitudes $\langle |\delta B_{y}|/|\mathbf{B}| \rangle = 0.003$ and $\langle |\delta B_{y}|/|\mathbf{B}| \rangle = 0.016$ respectively. In runs $SI$ and $SJ$ we drive fluctuations with frequency ($\omega_{0}/\omega_{pe} = 0.2$), with wavelength $\lambda = \Delta$ and with amplitudes $\langle |\delta B_{y}|/|\mathbf{B}| \rangle = 0.021$ and $\langle |\delta B_{y}|/|\mathbf{B}| \rangle = 0.136$ respectively. Table \ref{tab:com_cases} summarises the parameter space that we explore in this study.
%---------------------------------------------------------------------

%---------------------------------------------------------------------
\section{Simulation results}
\label{sec:simulation_results}

%---------------------------------------------------------------------
%In table \ref{tab:com_cases} we present the parameters that we use to perform the simulations in this study.
%--------------------------------------------------------------
%\begin{table}[h!]
\begin{table*} % <-- HERE
\begin{center}
\def~{\hphantom{0}}
\begin{tabular}{lccccccc}
\hline
$run$ & $SA$ & $SB$  & $SC$ & $SD$  & $SE$  & $SF$ & $SG$  \\[3pt]
\hline
$\frac{B_{g}}{B_{0}}$ & 0.1 & 0.1  & 0.1 & 0.1  & 0.2 & 0.2  & 0.1 \\
$\frac{n_{b}}{n_{0}}$ & 0.1 & 0.1  & 0.1 & 0.1  & 0.2  & 0.1  & 0.2 \\
$ \langle\frac{|\partial B_{y}| }{|\mathbf{B}|}\rangle$ & - & 0.010 & 0.002 & 0.004 & 0.016  & -  & 0.003 \\
$\frac{\omega_{0}}{\omega_{pe}}$ & - & - & 0.004 & 0.004 & 0.005  & -  & 0.2 \\
$\frac{\gamma_{0} }{\omega_{0}}$ & - & - & 0.6 & 0.6 & 0.9  & - & 0.6 \\
$N$-modes & - & 1 & 8 & 32 & 32  & -  & 32 \\
$m_{y} $ & - & 0.5 & 4 & 1,2,3,4 & 1,2,3,4 & -  & 1,2,3,4 \\
$m_{z} $ & - & 1 & 4 & 1,2,3,4 & 2,4,6,8 & - & 2,4,6,8  \\
\hline
$run$ & $SH$  & $SI$ & $SJ$ & $SK$ & $SL$ & $SM$ &  $SN$  \\[3pt]
\hline
$\frac{B_{g}}{B_{0}}$ & 0.1  & 0.1&  0.1  & 0.1 & 0.1  & 0.1&  0.1\\
$\frac{n_{b}}{n_{0}}$ & 0.2  & 0.2 &  0.2  & 0.2 & 0.2  & 0.2 &  0.2\\
$ \langle\frac{|\partial B_{y}| }{|\mathbf{B}|}\rangle$ & 0.016 & 0.021 &  0.136 & 0.069 & 0.036 & 0.069 & 0.036\\
$\frac{\omega_{0}}{\omega_{pe}}$ & 0.2 & 0.2 &  0.2 & 0.2 & 0.2 & 1 & 1 \\
$\frac{\gamma_{0} }{\omega_{0}}$ & 0.6 & 0.6 & 0.6 & 0.6 & 0.6 & 0.6 & 0.6 \\
$N$-modes & 32 & 8 & 8 & 8 & 8 & 8 & 8 \\
$m_{y} $ & 1,2,3,4 & 32  & 32 & 128 & 256 & 128 & 256\\
$m_{z} $ & 2,4,6,8 & 64 &  64 & 256 & 512 & 256 & 512\\
[3pt]
\hline
\end{tabular}
\caption{List of parameters that we explore in this study.}
\label{tab:com_cases}
\end{center}
%\end{table} 
\end{table*} % <-- HERE
%--------------------------------------------------------------

\begin{figure*}{}
\centering
\begin{tikzpicture}
\draw (-7.9, 5.7) node[inner sep=0] {\includegraphics[width=0.34\linewidth]{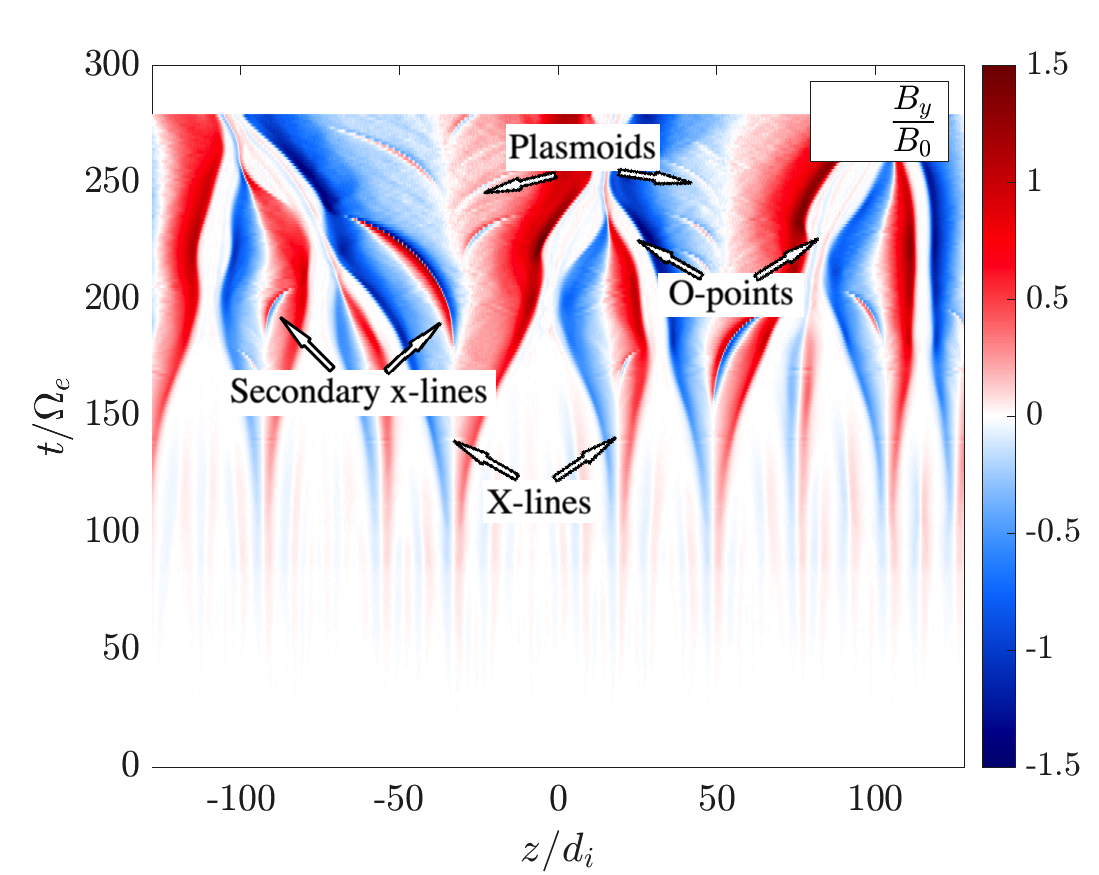}};
\draw (-1.9, 5.7) node[inner sep=0] {\includegraphics[width=0.34\linewidth]{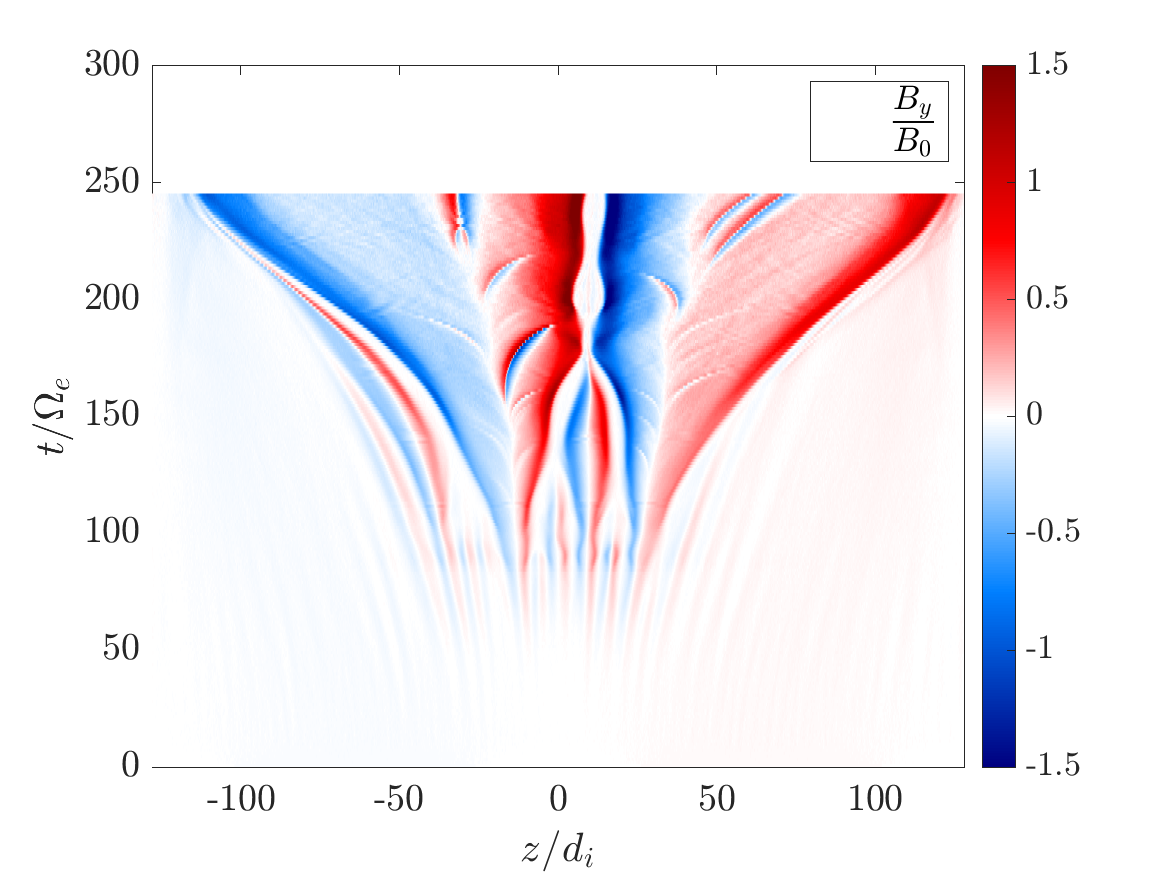}};
\draw (4, 5.7) node[inner sep=0] {\includegraphics[width=0.34\linewidth]{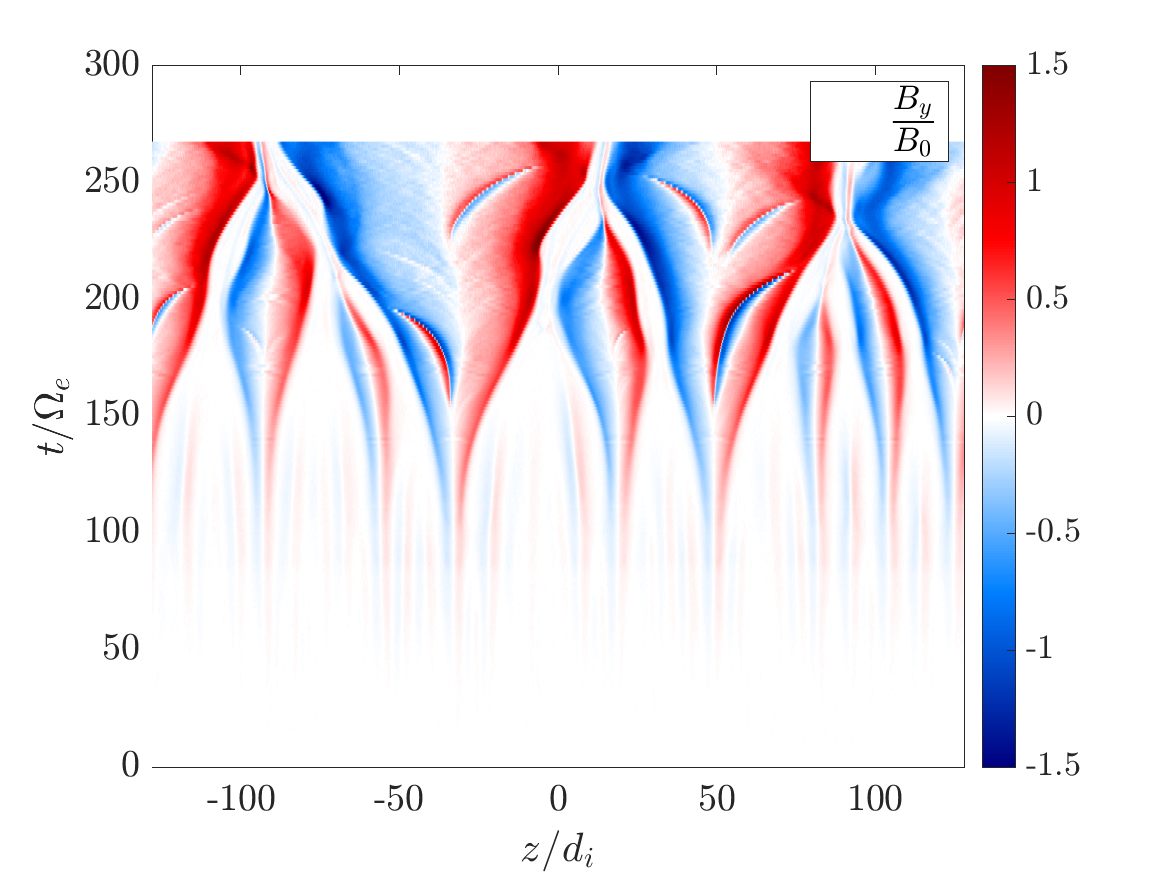}};
\draw (-7.9, 1) node[inner sep=0] {\includegraphics[width=0.34\linewidth]{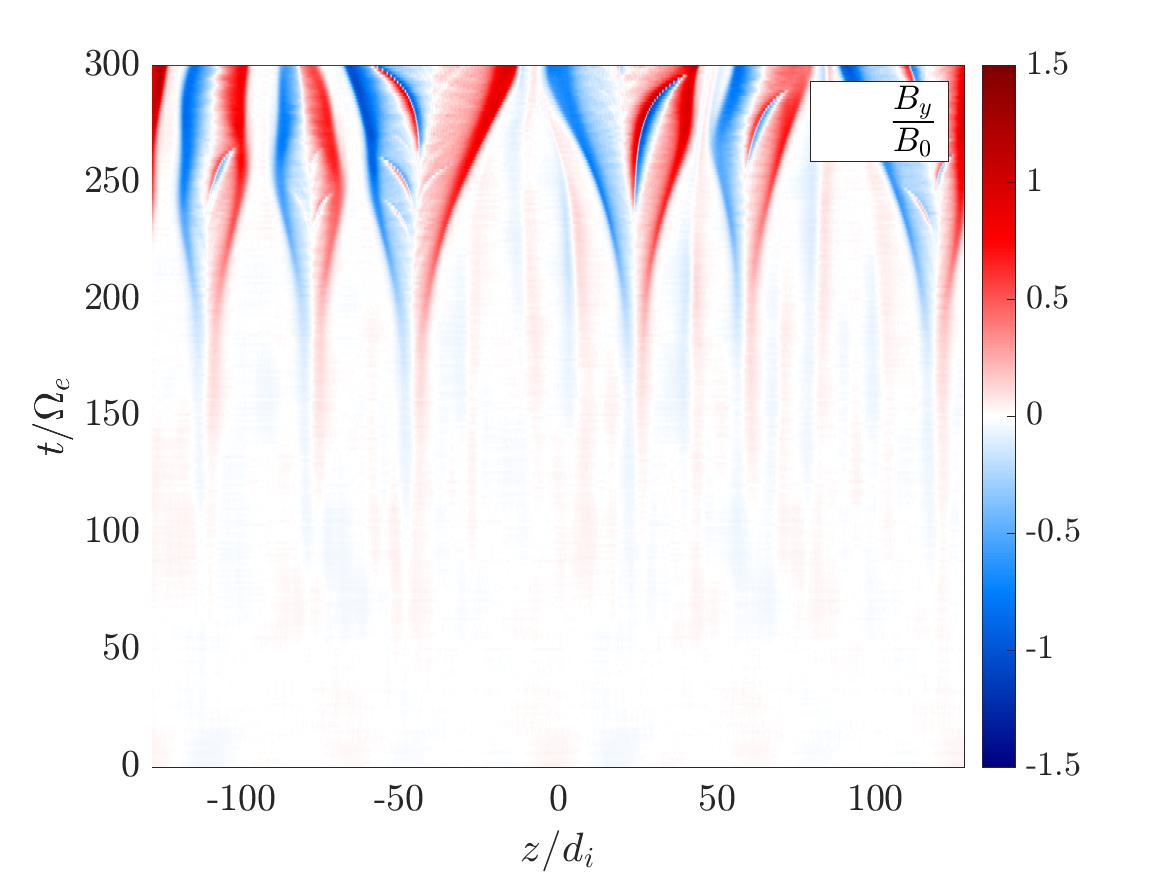}};
\draw (-1.9, 1) node[inner sep=0] {\includegraphics[width=0.34\linewidth]{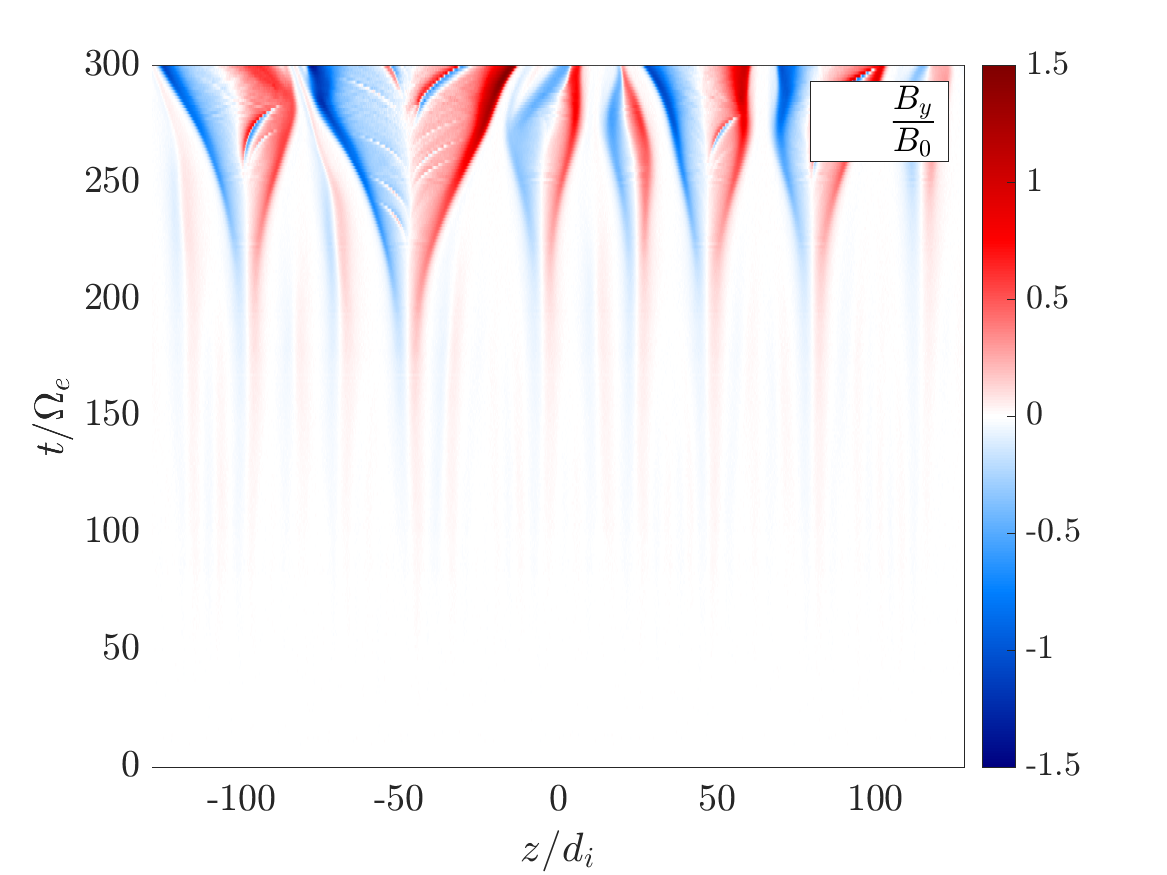}};
\draw (4, 1) node[inner sep=0] {\includegraphics[width=0.34\linewidth]{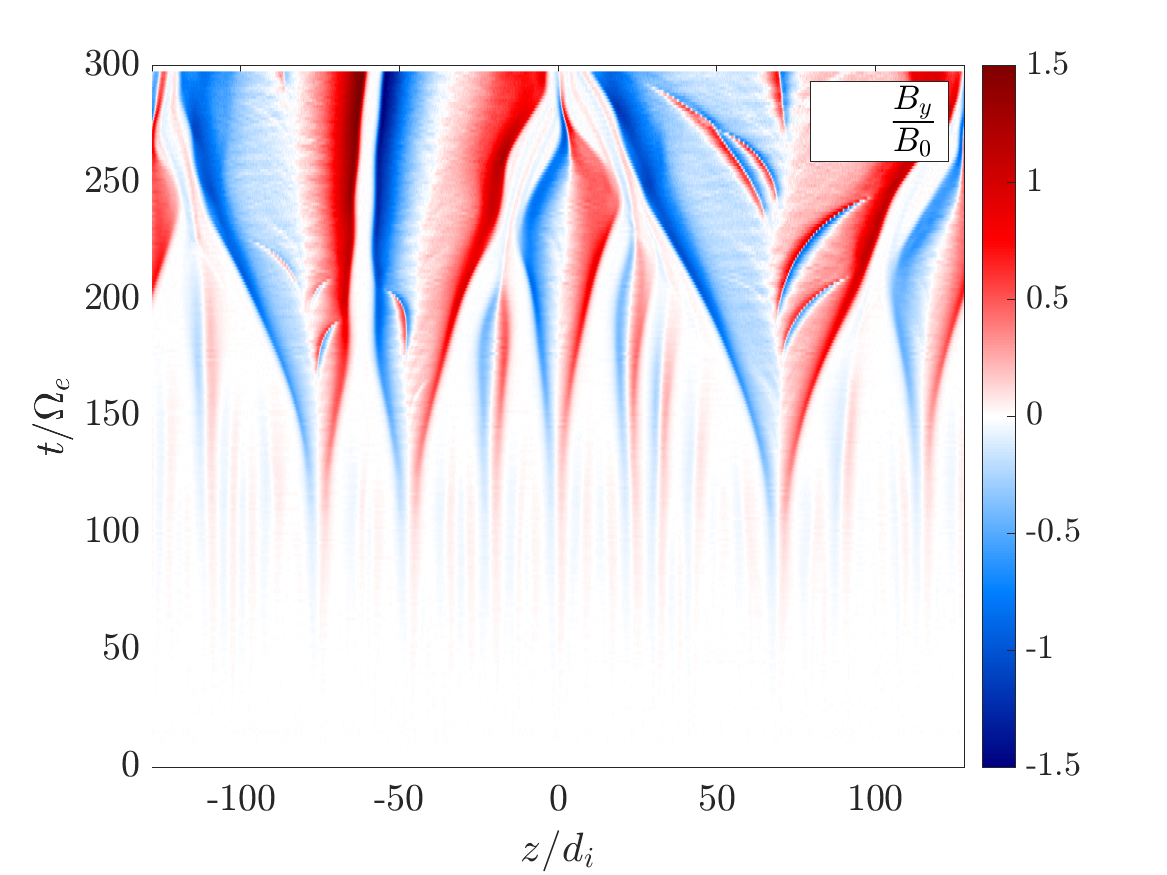}};
\draw (-7.9, -3.7) node[inner sep=0] {\includegraphics[width=0.34\linewidth]{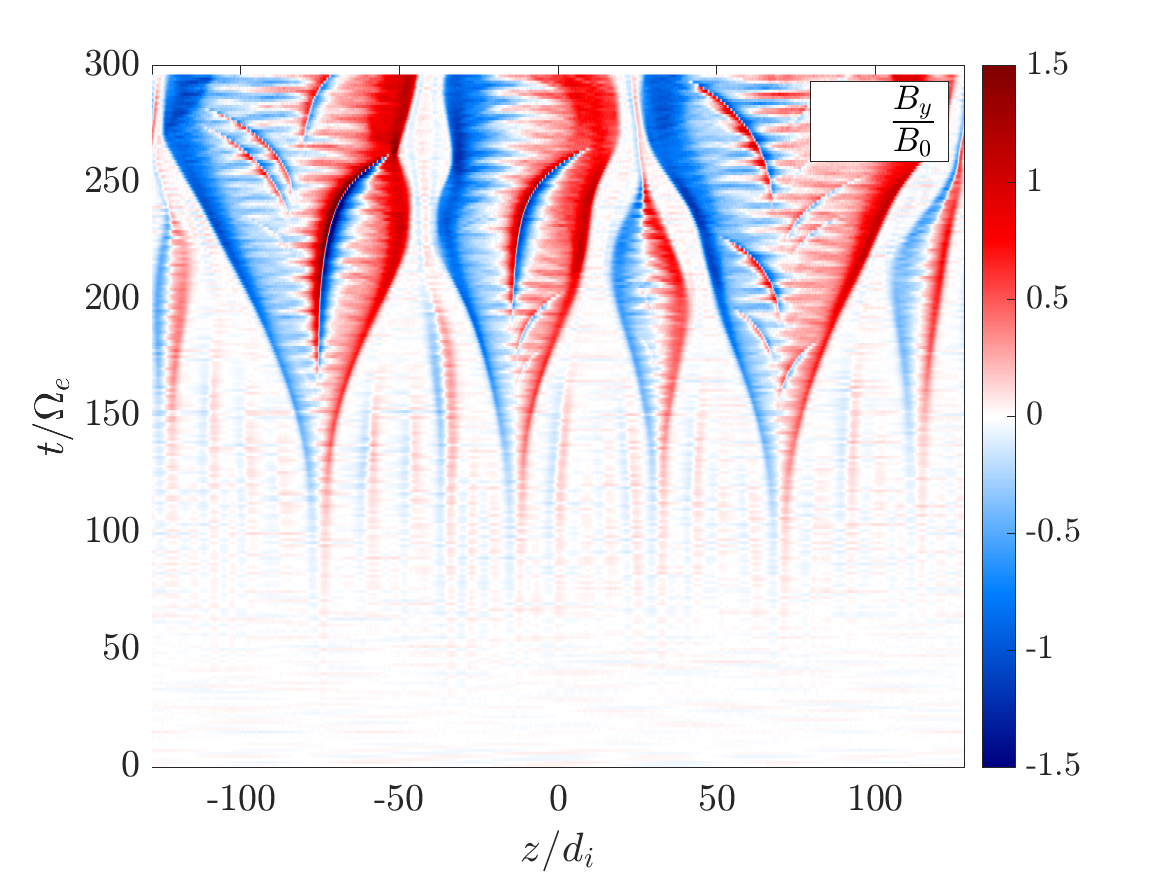}};
\draw (-1.9, -3.7) node[inner sep=0] {\includegraphics[width=0.34\linewidth]{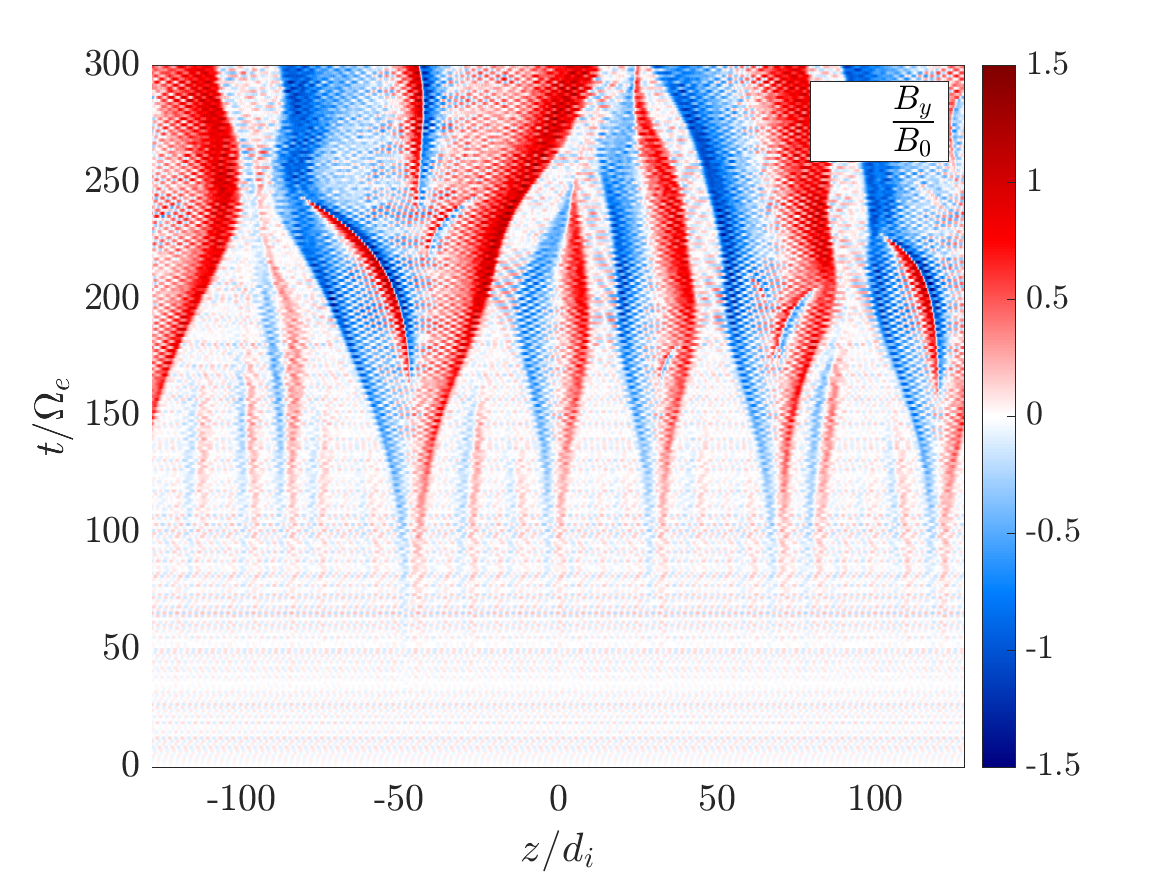}};
\draw (4, -3.7) node[inner sep=0] {\includegraphics[width=0.34\linewidth]{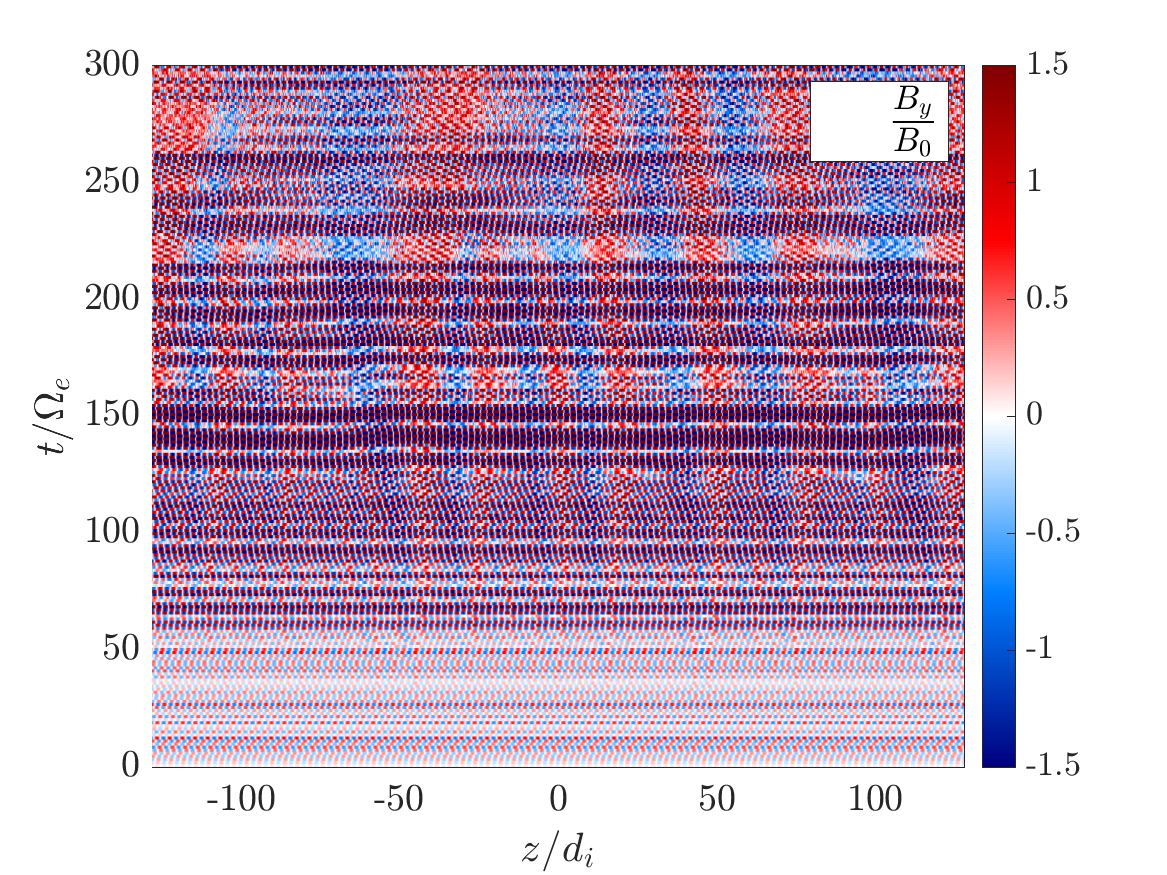}};
\draw (-10.5, 8.0) node {(a) SA};
\draw (-4.7, 8.0) node {(b) SB};
\draw (1.2, 8.0) node {(c) SC};   
\draw (-10.5, 3.2) node {(d) SE};
\draw (-4.7, 3.2) node {(e) SF};
\draw (1.2, 3.2) node {(f) SG};   
\draw (-10.5, -1.4) node {(g) SH};
\draw (-4.7, -1.4) node {(h) SI};
\draw (1.2, -1.4) node {(i) SJ};  
\end{tikzpicture}
\caption{Stack plots of 1D cuts of the reconnecting magnetic-field component $B_{y}$ taken along the line $y=0$ for runs: (a) $SA$, (b) $SB$ , (c) $SC$, (d) $SE$, (e) $SF$, (f) $SG$, (g) $SH$, (h) $SI$ and (i) $SJ$.  {Note that we stop runs $SA$, $SB$ and $SC$ at an earlier time step than runs $SD-SI$ as the formation of growing magnetic islands and their time evolution is already visible.} 
} 
\label{fig:By_stack}
\end{figure*}
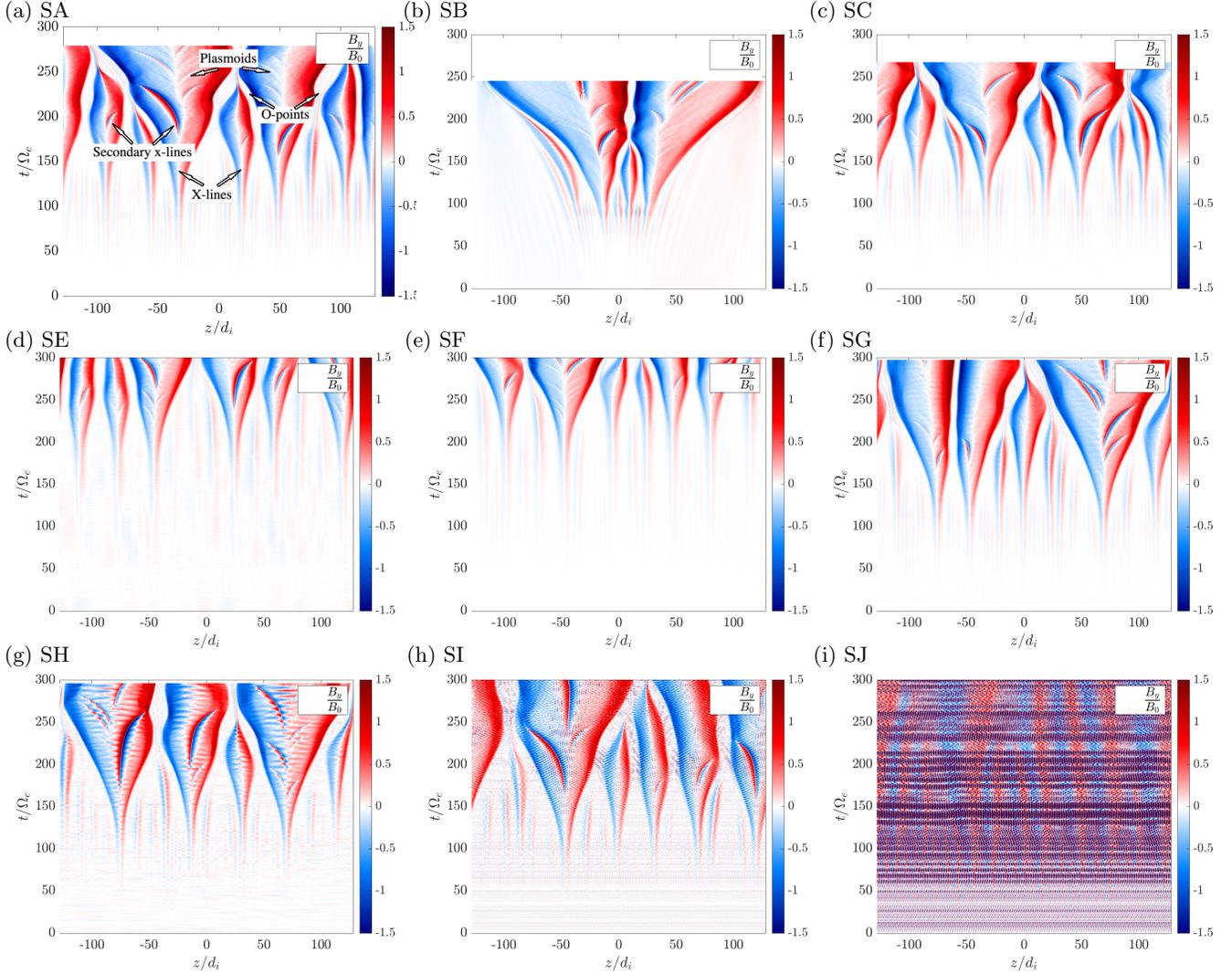

%---------------------------------------------------
In this section we describe the effect of the driven fluctuations on the onset of the tearing instability. To visualise the time evolution of the system, we take 1D cuts of $B_{y}$ at $y=0$ at each time step and construct a stack plot. Figure \ref{fig:By_stack} depicts stack plots for a selection of the aforementioned runs. The blue (red) colour represent negative (positive) values of the $B_{y}$. The growing blue-red patterns show the formation of x-points and extended current sheets between magnetic islands due to the tearing instability. 

Panel \ref{fig:By_stack}a) shows the evolution of run $SA$. At $t \sim 60 \ \Omega_{e}^{-1}$ there is a well defined signature of plasmoids formation. At $t \sim 150 \ \Omega_{e}^{-1}$ there are 8 short reconnecting current-sheets. At $t \sim 260 \ \Omega_{e}^{-1}$ there are only three reconnecting current-sheets and three o-points. The white stripes within the current-sheets, e.g, at $z=50 d_{i}$ and between $t\sim 150$ to $t \sim 200 \ \Omega_{e}$ correspond to drifting plasmoids that are expelled out of the current-sheet and adsorbed by the growing magnetic islands.  

Panel \ref{fig:By_stack}b) shows the evolution of run $SB$. Similar to run $SA$, by $t \sim 60 \ \Omega_{e}^{-1}$ there is ongoing formation of plasmoids. However, unlike run $SA$, run $SB$ at $t \sim 150 \ \Omega_{e}^{-1}$ there are only 4 reconnecting currents and as early as $t \sim 200 \ \Omega_{e}^{-1}$ there are only two extended currents sheets and two main o-points.  

Panel \ref{fig:By_stack}c) shows the evolution of run $SC$. In this run, although the driven fluctuations modify the background plasma, the local dynamics of plasmoids formation is dominant and the system evolves similar to run $SA$ with 8 reconnecting current sheets at $t \sim 150 \ \Omega_{e}^{-1}$ and only three at $t \sim 250 \ \Omega_{e}^{-1}$. The same is true for run $SD$ (not shown here). Panels \ref{fig:By_stack}d) and \ref{fig:By_stack}e) depict the evolution of runs $SE$ and $SF$ respectively. In both runs the formation of plasmoids is delayed to $t \sim 100 \ \Omega_{e}^{-1}$ showing the effect of a larger guide field. For run $SE$ there are 6 reconnecting current-sheets by $t \sim 250 \ \Omega_{e}^{-1}$ whereas for run $SF$ there are 7 reconnecting current-sheets by $t \sim 250 \ \Omega_{e}^{-1}$. 

Panels \ref{fig:By_stack}f), \ref{fig:By_stack}g) and \ref{fig:By_stack}h) show the system evolution for runs $SG$, $SH$ and $SI$ respectively. For these runs the driven fluctuations are visible within the reconnecting currents-sheets and the overall evolution is similar to run $SA$. 

Panel \ref{fig:By_stack}i) shows the stack plot for run $SJ$, in spite of the current-sheet breaking down into smaller segments, there is no formation of o-points and the magnetic islands do not grow similar to run $SA$. We refer to this process as suppression of the tearing mode. 

We further explore the effect of driving high-frequency and short-wavelength fluctuations. In runs $SK$ and $SL$ we drive eight waves with $\omega_{0} = \Omega_{e}$ and wavelengths $\lambda = d_{e}$ and $\lambda = \rho_{e}$ respectively. In runs $SM$ and $SN$ we drive eight waves with $\omega_{0} = \omega_{pe}$ and wavelengths $\lambda = d_{e}$ and $\lambda = \rho_{e}$ respectively. We observe that driving fluctuations at kinetic scales with small amplitudes ($|\delta B_{y}|/B_{0}<0.069$) (not shown here) prevents the tearing instability from growing similar to run $SJ$. Moreover, for these 5 runs there is no evidence of extended current-sheets nor drifting plasmoids.

\begin{figure*}{}
\centering
\begin{tikzpicture}
\draw (-5.3, 5.7) node[inner sep=0] {\includegraphics[width=1.0\linewidth]{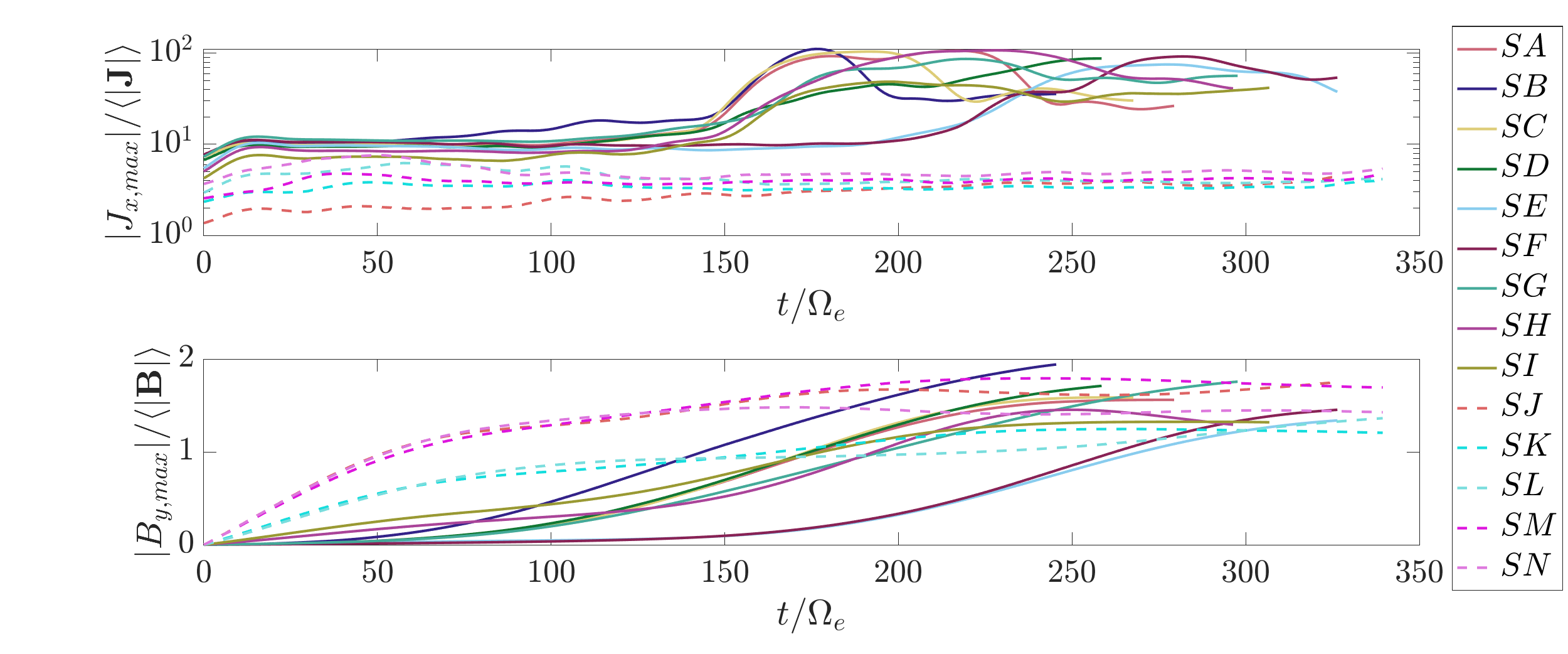}};
\draw (-13.5, 9.0) node {a)};
\draw (-13.5, 5.2) node {b)};   
\end{tikzpicture}
\caption{Time evolution of a) the absolute value of the maximum out-of-plane current density $|J_{x,max}|$ and b) the absolute value of the reconnecting magnetic field component $|B_{y,max}|$ where we substract its initial value ($|B_{y,max}|_{t=0}$) to facilitate the comparison between runs. The solid-lines correspond to runs in which the magnetic islands grow regardless of the driven fluctuations. Dashed-lines correspond to runs in which the growing of the magnetic islands is suppressed.}
\label{fig:currents_max}
\end{figure*}

As the reconnection takes place, the out-of-plane current grows. Figure \ref{fig:currents_max}a) shows the absolute value of the maximum out-of-plane current density $|J_{x,max}|$ normalized to the spatial average $\langle |\vec{J}| \rangle$. The solid-lines correspond to runs in which the magnetic islands grow. For runs $SA, \ SB, \ SC, \ SD, \ SG, \ SH$ and $SI$, there is an exponential grow at $t \sim 150 \ \Omega_{e}$. For run $SB$ there is a peak at $t \sim 180 \ \Omega_{e}$ whereas for the other runs there is an extended plateau. The exponential grow for runs $SE$ and $SF$ starts at $t \sim 200 \ \Omega_{e}$ which appears to be correlated to the higher guide magnetic field. The dashed-lines correspond to runs with non-growing magnetic islands ($SJ, \ SK, \ SL, \ SM$ and $SN$). For these runs unlike for the previous runs, $|J_{x,max}|$ remains approximately constant in time showing no exponential grow.  

To further explore the effect of driving fluctuations on the reconnection dynamics, we study the time evolution of $B_{y,max}$ as a proxy of the fastest growing mode. Figure \ref{fig:currents_max}b) shows the absolute value of the maximum reconnecting magnetic-field component $|B_{y,max}|$. The solid-lines correspond once more to growing magnetic-islands runs ($SA$ to $SI$). For these runs there is an initial slow grow of $|B_{y,max}|$ followed by a faster grow that saturates at later time. For these runs the saturation plateau corresponds to $ |B_{y,max}| \sim (1.6 - 1.8) \ \langle |\mathbf{B}| \rangle$, except for run $SB$ that at its latest time peaks at $ |B_{y,max}| \sim 1.9 \ \langle |\mathbf{B}| \rangle$.
 
The dashed-lines correspond to non growing magnetic-islands runs ($SJ$ to $SN$). For these runs the time evolution of $|B_{y,max}|$ follows an initial grow faster than their growing magnetic island counterpart (solid-lines), followed by a slow grow towards a plateau in the same range $ |B_{y,max}| \sim (1.6 - 1.8) \ \langle |\mathbf{B}| \rangle$.  

Assuming that grows as $B_{y}(t) = B_{y,0}e^{\gamma_{By}t}$, where $B_{y,0}$ is an initial value and $\gamma_{By}$ is the growth rate, we estimate the growth rate as 

\begin{align}
    \gamma_{By} = \ln \left( \frac{|B_{y}(t + \Delta t)|}{|B_{y}(t)|} \right) \frac{1}{\Delta t},
    \label{eqn:gamma}
\end{align}

\noindent where ${\Delta t}^{-1} = 0.766 \ \Omega_{e}$ is the output interval. {Note that the fast and short grow of $|B_{y,max}|$ for runs $SJ-SN$ do not correspond to the growth of magnetic islands but it reflects the dominant nature of the injected fluctuations in these runs. To account for this difference we compute the time average $\langle \gamma_{By} \rangle_{t}$ over the entire simulation time. This measure better captures the effect of fluctuations on the growth rate associated to the growing of magnetic islands.} Table \ref{tab:com_gammas} lists the time average $\langle \gamma_{By} \rangle_{t}$ of the growth rate $\gamma_{By}$ and its maximum value $\gamma_{By,max}$. For all the runs in which the magnetic islands grow ($SA$ to $SI$), $\{ \langle \gamma_{By} \rangle_{t}/\Omega_{e},\max (\gamma_{By})/\Omega_{e} \} > 7\times 10^{-3}$, whereas for all non-growing runs $\{\langle \gamma_{By} \rangle_{t}/\Omega_{e},\max (\gamma_{By})/\Omega_{e}\} < 5 \times 10^{-3}$.    

%---------------------------------------------------------------------------------------------------------------------------------------------------------------

%%-------------------------------------------------------------------
%%-------------------------------------------------------------------

%-------------------------------------------------------------------
%-------------------------------------------------------------------

%--------------------------------------------------------------
%\begin{table}[h!]
\begin{table*} % <-- HERE
\begin{center}
\def~{\hphantom{0}}
\begin{tabular}{lcccccccccccccc}
\hline
$run$ & $SA$ & $SB$  & $SC$ & $SD$  & $SE$  & $SF$ & $SG$ & $SH$  & $SI$ & $SJ$ & $SK$ & $SL$ & $SM$ &  $SN$  \\[3pt]
\hline
$(\langle \gamma_{By} \rangle_{t} / \Omega_{e}) 10^{-2}   $ & 1.8 & 1.7 & 1.8 & 1.7 & 1.0 & 1.6 & 1.5 & 0.8 & 0.7 &  0.3 & 0.4 & 0.5 & 0.4 & 0.3\\
$(max(\gamma_{By}) / \Omega_{e}) 10^{-2} $ & 3.0 & 3.6 & 3.0 & 2.9 & 2.1 & 2.4 & 2.6 & 1.3 & 0.9 &  0.2 & 0.2 & 0.2 & 0.2 & 0.1\\
[3pt]
\hline
\end{tabular}
\caption{Time average $\langle \gamma_{By} \rangle_{t}$ and maximum $max(\gamma_{By})$ of the growth rate $\gamma_{By}$ (Eq. \ref{eqn:gamma}) for the runs we explore in this study.}
\label{tab:com_gammas}
\end{center}
%\end{table} 
\end{table*} % <-- HERE
%--------------------------------------------------------------

%-------------------------------------------------------------------
    \begin{figure*}
        \centering
        \includegraphics[width=1.0\textwidth]{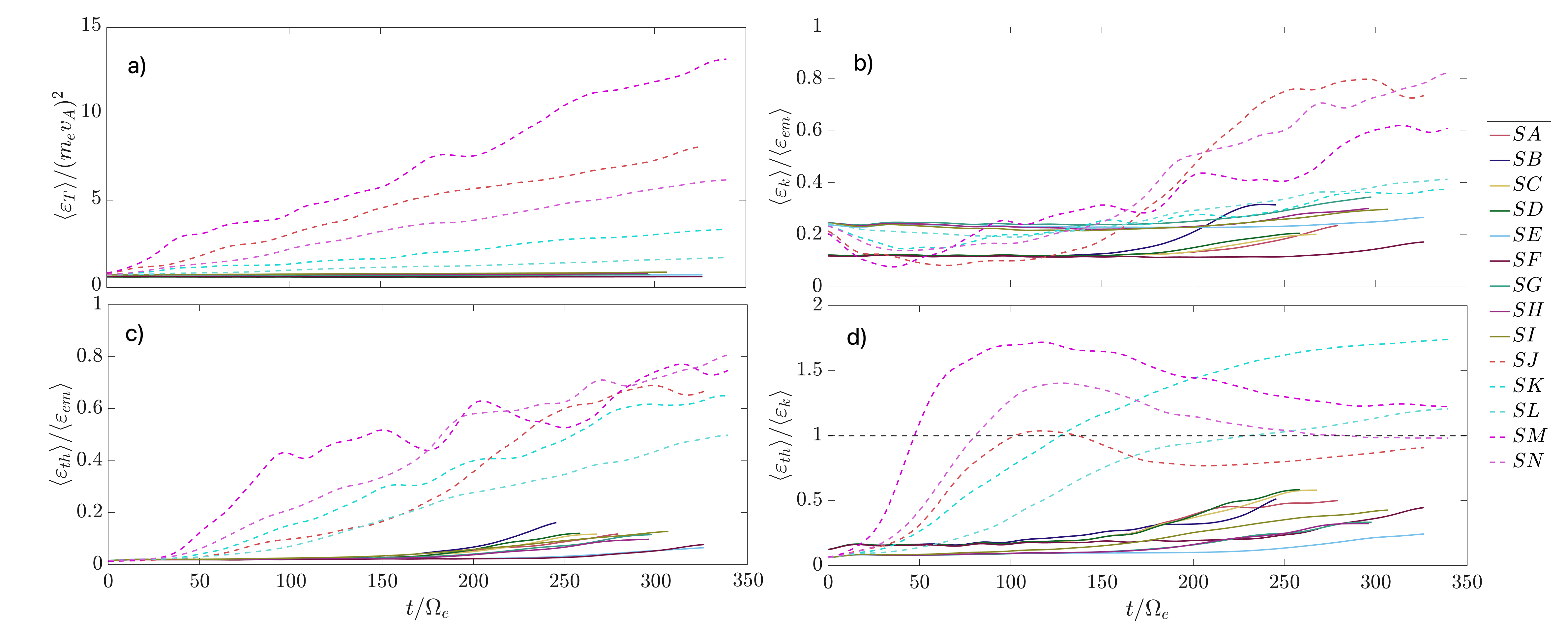}
        \caption{a) Average total energy density budget vs time; b) electron kinetic to electromagnetic energy density ratio $\langle \varepsilon_{k,e}\rangle / \langle \varepsilon_{em} \rangle$; c) electron thermal to electromagnetic energy densities ratio $\langle \varepsilon_{th,e} \rangle / \langle \varepsilon_{em} \rangle$; d) electron thermal to electron kinetic energy densities ratio $\langle \varepsilon_{th,e} \rangle / \langle \varepsilon_{k,e} \rangle$. The solid-lines correspond to growing magnetic-islands runs. The dashed-lines correspond to non-growing magnetic islands runs.}
        \label{fig:enterlabel12}
    \end{figure*}

       \begin{figure*}
        \centering
        \includegraphics[width=1.0\textwidth]{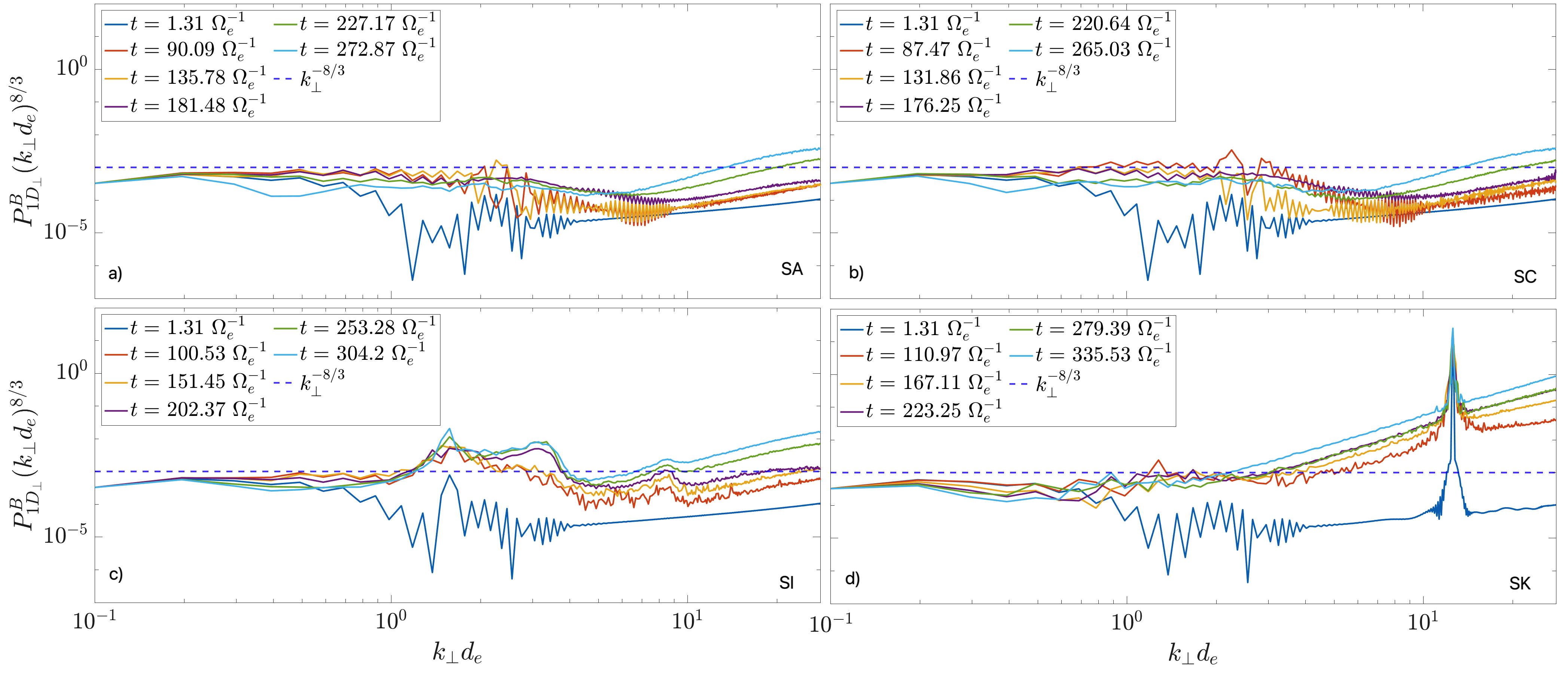}
        \caption{Compensated 1D power spectrum of the magnetic fluctuations, $P_{1D_{\perp}}^{B}(k_{\perp}d_{e})^{8/3}$, as function of $k_{\perp}$ for: a) run $SA$, b) run $SC$, c) run $SI$ and d) run $SK$. The curves are colour-coded with the time step at which the spectrum is computed. The progressive order in time is Blue (t1), orange (t2), yellow (t3), purple (t4), green (t5) and cyan (t6).}
        \label{fig:enterlabel13}
    \end{figure*}
%-------------------------------------------------------------------

%-------------------------------------------------------------------
\begin{figure}{}
\centering
\includegraphics[scale=0.48]{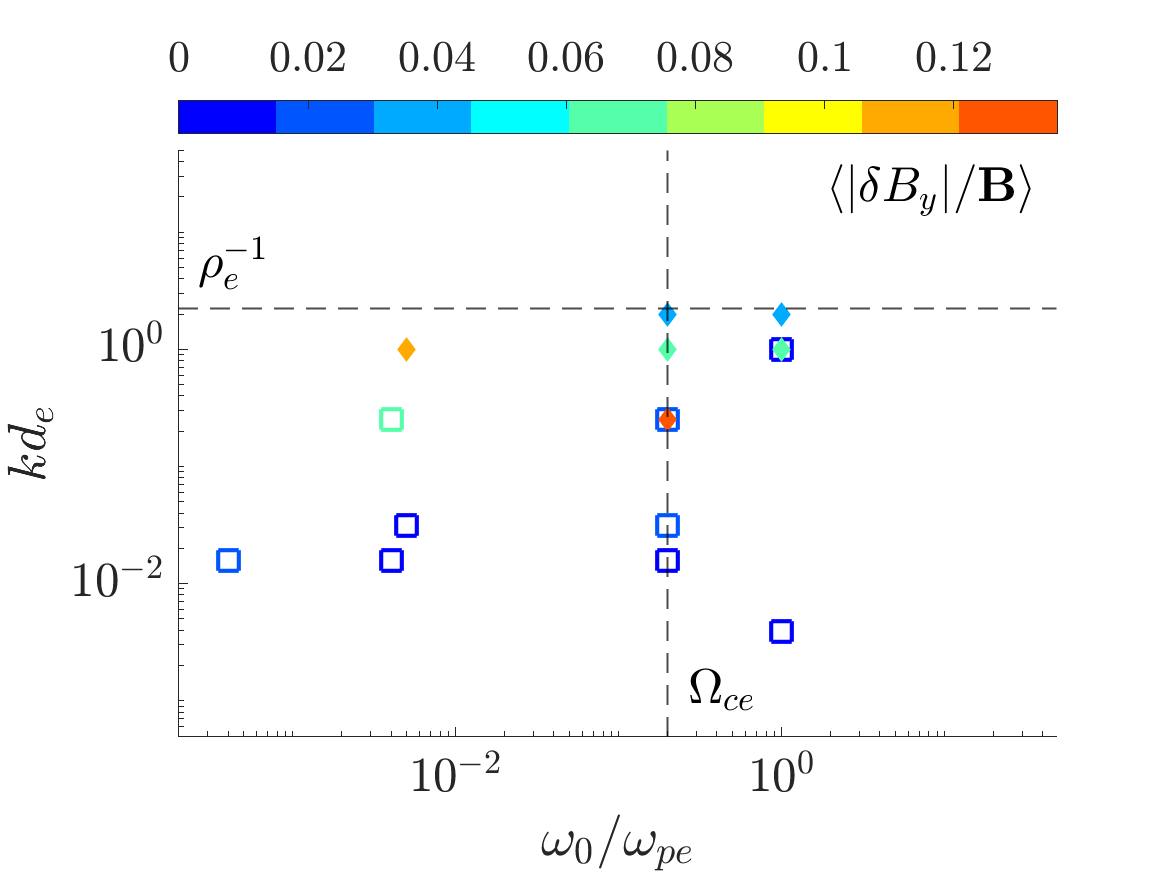}
\caption{Parameter space explored in this study. $\omega_{0}$ and $k =  2\pi/\lambda$ are the frequency and wavenumber of  the driven fluctuations respectively. The data points are color-coded with the amplitude of the fluctuations $\langle |\delta B_{y}|/|\mathbf{B}|\rangle$. The empty squares represent runs in which the magnetic islands grow. The solid diamonds represent runs for which the magnetic islands do not grow.} 
%The gray areas shows the regions where the tearing instability is suppressed by the driven fluctuations with a given amplitude.
\label{fig:parameterspace}
\end{figure}
%-------------------------------------------------------------------

%-------------------------------------------------------------------
\begin{figure*}{}
\centering
\includegraphics[scale=0.5]{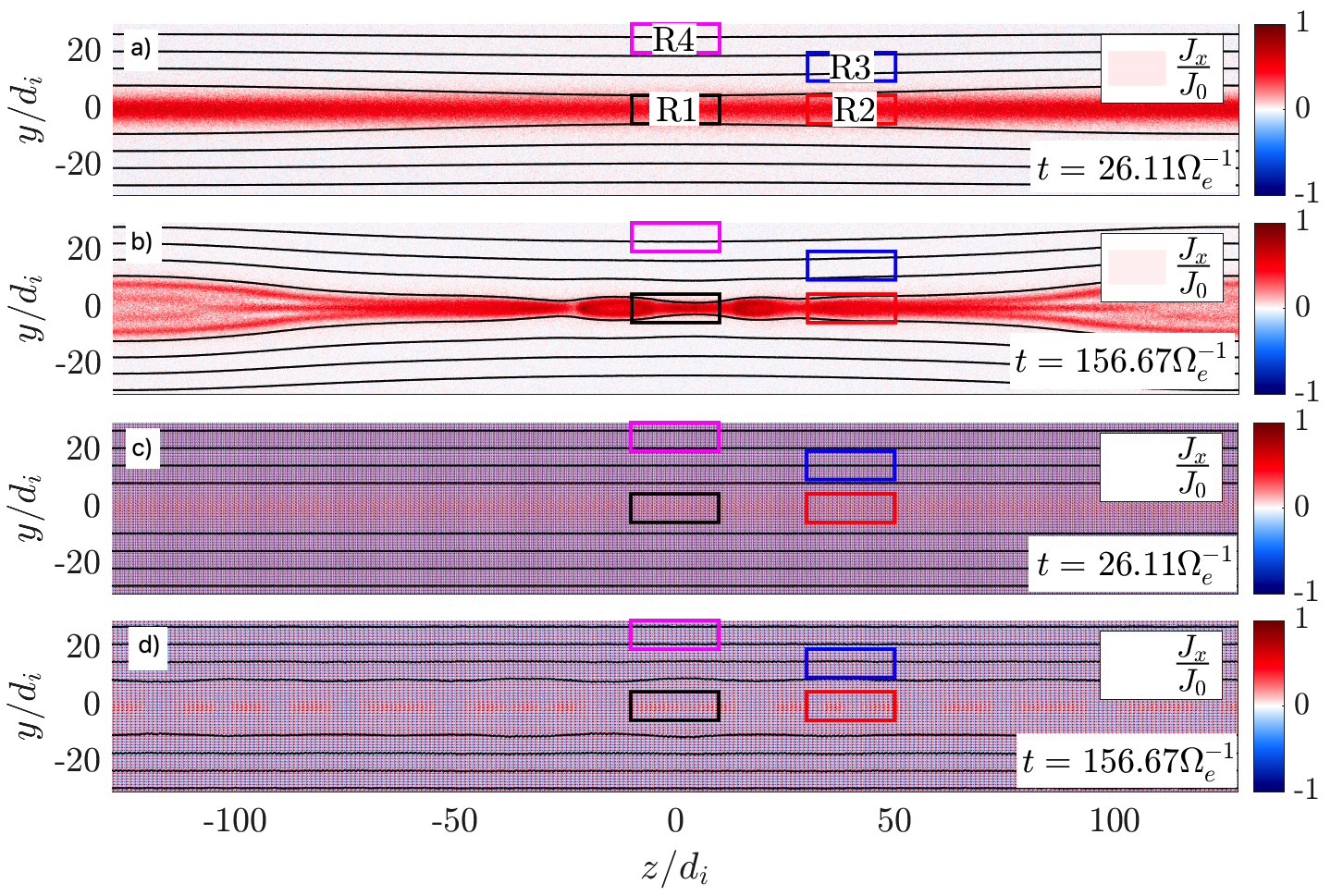}
\hfill
\caption{Out-of-plane electric current density $J_{x}$ normalized to $J_{0}=max(|J_{x}|)$ for: a) run $SB$ at $t=26.11 \Omega_{e}^{-1}$, b) run $SB$ at $t=156.67 \Omega_{e}^{-1}$, c) run $SJ$ at $t=26.11 \Omega_{e}^{-1}$ and d) run $SJ$ at $t=156.67 \Omega_{e}^{-1}$. The squares highlight regions where we analyze the electron velocity distribution  function (eVDF).}
\label{fig:currentdensities}
\end{figure*}
%-------------------------------------------------------------------

%-------------------------------------------------------------------
\begin{figure*}{}
\centering
\includegraphics[width=1.0\textwidth]{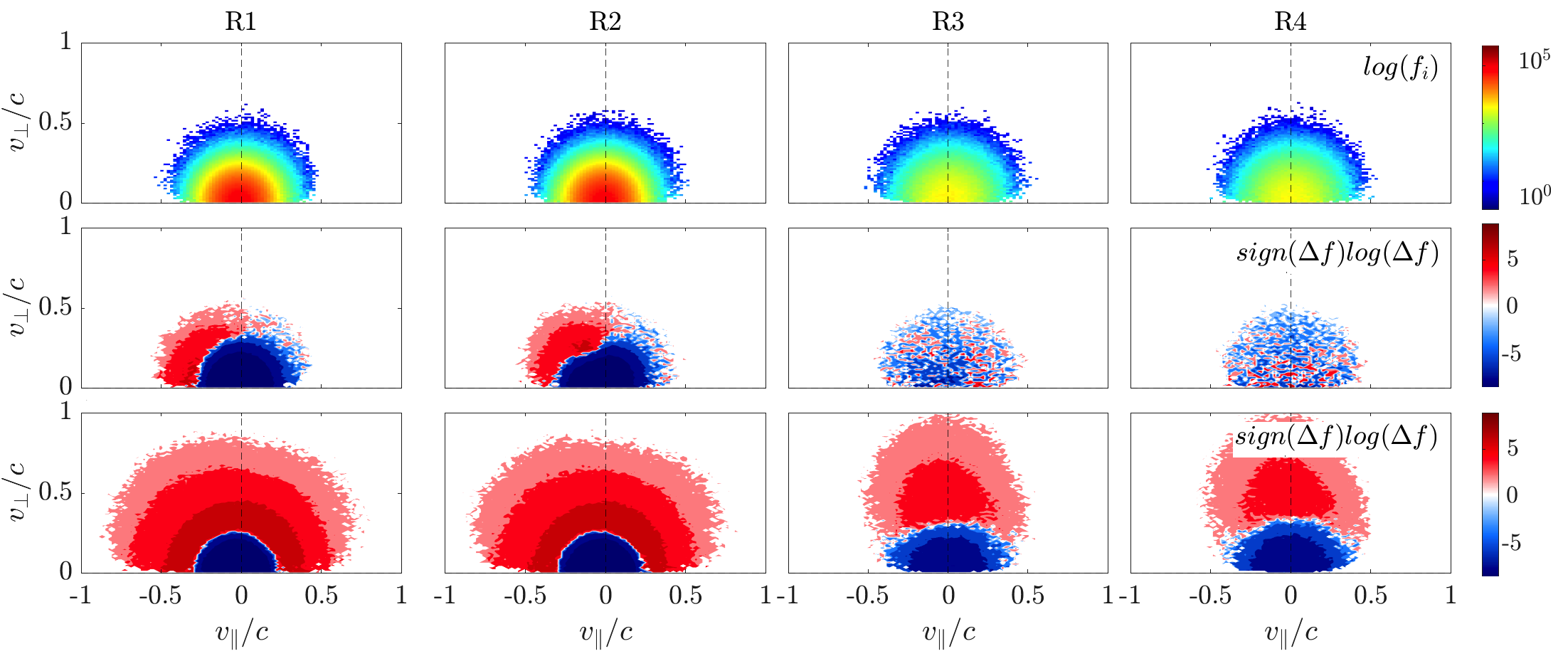}
\caption{Top-row: electron VDF in the $v_{\parallel}-v_{\perp}$ plane for run $SB$ at $t=26.11 \Omega_{e}^{-1}$. The columns $R1$, $R2$, $R3$ and $R4$ correspond to the regions highlighted with the squares in Figure \ref{fig:currentdensities}. Middle-row: electron VDF difference ($\Delta f = f_{t2} - f_{t1}$) for run $SB$ where $t1=26.11 \Omega_{e}^{-1}$ and $t2=156.67 \Omega_{e}^{-1}$. Bottom-row: same as middle-row for run $SJ$.}
\label{fig:VDFs}
\end{figure*}
%---------------------------------------------------------------------

%---------------------------------------------------------------------------------------------------------------------------------------------------------------
\subsection{Energy and spectral time evolution analysis}

{To address the question about the energy conservation, we define the total energy density budget as $\langle \varepsilon_{T} \rangle = \langle \varepsilon_{em} \rangle + 2 \langle \varepsilon_{k,e} \rangle + 2 \langle \varepsilon_{th,e} \rangle$, where $\langle \ldots \rangle$ is the average over the simulation domain, $\varepsilon_{em} = (|B|^{2}/\mu_{0} + \varepsilon_{0}|E|^2)/2$ is the electromagnetic energy density, $\varepsilon_{k,e} = n_{e}m_{e}|u_{e}|^{2}/2$ is the electron kinetic energy density, $\varepsilon_{th,e} = Tr(\overline{{\mathbf{P}}}_{e})/2$ is the electron thermal energy density and $\overline{{\mathbf{P}}}_{e}$ is the electron pressure tensor $\overline{{\mathbf{P}}}_{e} = m_{e} \int f(\mathbf{v})(\mathbf{u}_{e}-\mathbf{v})(\mathbf{u}_{e} - \mathbf{v})d^{3}v$ where $f(\mathbf{v})$ is the electron velocity distribution function.}
    
{Figure (\ref{fig:enterlabel12}a) shows $\langle \varepsilon_{T} \rangle$ as a function of time for the different runs. The solid-lines represent growing magnetic-islands runs and the dashed-lines represent non-growing magnetic-islands runs. The colour code is the same a figure \ref{fig:currents_max}.  For run $SA$ and $SB$ the total energy budget remains approximately constant. For runs $SC-SI$ the increase of the energy density budget due to the fluctuations that we drive into the system is negligible. In contrast, for the non-growing magnetic island runs (dashed-lines), the sustained increase in $\langle \varepsilon_{T} \rangle$ shows the accumulation of energy in the simulation. Although, the energy is not conserved due to the injection of energy, it is worth to explore the exchange of energy balance. Thus, we use the ratio between the electron kinetic and electromagnetic energy densities $\langle \varepsilon_{k,e}\rangle / \langle \varepsilon_{em} \rangle$, the ratio between the electron thermal and electromagnetic energy densities $\langle \varepsilon_{th,e} \rangle / \langle \varepsilon_{em} \rangle$, and the ratio between the electron thermal and electron kinetic energy densities $\langle \varepsilon_{th,e} \rangle / \langle \varepsilon_{k,e} \rangle$.}  
    
{Figure (\ref{fig:enterlabel12}b) shows the $\langle \varepsilon_{k,e}\rangle / \langle \varepsilon_{em} \rangle$ for the different runs. For all runs, $\langle \varepsilon_{k,e}\rangle < \langle \varepsilon_{em} \rangle$ at all times. For runs $SA-SI$, $\langle \varepsilon_{k,e}\rangle / \langle \varepsilon_{em} \rangle$ remains approximately constant until the magnetic-island growing begins (see Figure \ref{fig:By_stack}) and the particles gain kinetic energy. For run $SB$, the ratio reaches the plateau $\langle \varepsilon_{k,e}\rangle / \langle \varepsilon_{em} \rangle = 0.3$. In contrast for runs $SJ-SN$, $\langle \varepsilon_{k,e}\rangle / \langle \varepsilon_{em} \rangle$ initially decreases due to the injection of the electromagnetic energy, and then increases showing that particles gain kinetic energy density at expenses of the magnetic energy density without the formation of magnetic-islands.}   
       
{Figure (\ref{fig:enterlabel12}c) shows the ratio $\langle \varepsilon_{th,e}\rangle / \langle \varepsilon_{em} \rangle$. For run $SA-SI$, $\langle \varepsilon_{th,e}\rangle / \langle \varepsilon_{em} \rangle$ remains approximately constant until the growing of the magnetic-islands starts similarly to $\langle \varepsilon_{k,e}\rangle / \langle \varepsilon_{em} \rangle$. In contrast, for runs $SJ-SN$, $\langle \varepsilon_{th,e} \rangle / \langle \varepsilon_{em} \rangle$ remains approximately constant for a short interval. This suggests that particles are gaining thermal energy as the electromagnetic energy is injected into the system. Afterwards, $\langle \varepsilon_{th,e}\rangle / \langle \varepsilon_{em} \rangle$ considerably increases.}

{Figure (\ref{fig:enterlabel12}d) shows the ratio $\langle \varepsilon_{th,e} \rangle / \langle \varepsilon_{k,e} \rangle$. The horizontal black dashed-line marks a ratio equal to one. For runs $SA-SI$, $\langle \varepsilon_{th,e} \rangle / \langle \varepsilon_{k,e} \rangle$ is approximately constant up to $t \approx 100 \Omega_{e}^{-1}$ and then $\langle \varepsilon_{th,e} \rangle / \langle \varepsilon_{k,e} \rangle$ slowly increases. Nonetheless, $\langle \varepsilon_{th,e} \rangle < \langle \varepsilon_{k,e} \rangle$ at all times. For runs $SJ, SM$ and $SN$, $\langle \varepsilon_{th,e} \rangle / \langle \varepsilon_{k,e} \rangle$ grows to a saturation value around one. In contrast, for runs $SK$ and $SL$ $\langle \varepsilon_{th,e} \rangle / \langle \varepsilon_{k,e} \rangle$ continuously increase.}  

{Previous studies have explored the turbulent nature of magnetic reconnection as well as the associated cross-scale energy transfer \citep{adhikari2020reconnection,adhikari2024scale}. To understand the effect of the injected fluctuations on the energy transport across the scales, we compute the perpendicular 1D power spectrum of the magnetic-field fluctuations $P^{B}_{1D_{\perp}}$ \citep{franci2018solar}, where we consider $\hat{y}$ ($\hat{z}$) the perpendicular (parallel) direction. Since we impose conductive boundary conditions along $\hat{y}$, we apply a hamming window \citep{oppenheim1999discrete} to enforce periodicity. Figure \ref{fig:enterlabel13} shows the compensated 1D power spectrum of the magnetic field fluctuations $P^{B}_{1D_{\perp}}(k_{\perp}d_{e})^{8/3}$ at six different times t1 (Blue), t2 (orange), t3 (yellow), t4 (purple), t5 (green) and t6 (cyan) for run $SA$ (a), run $SC$ (b), run $SI$ (c) and run $SK$ (d). We normalise $P^{B}_{1D_{\perp}}$ to its maximum value at each time step and the blue dashed-line represents a spectral slope $K_{\perp}-8/3$ which is expected for turbulence in magnetised pair-plasma \citep{loureiro2018turbulence}.}
     
{Figure (\ref{fig:enterlabel13}a) shows $P^{B}_{1D_{\perp}} (k_{\perp}d_{e})^{8/3}$ for run $SA$.  At the earliest time $t=1.31 \Omega_{e}^{-1}$, the spectrum of the current sheet structure corresponds to a superposition of modes. As the system evolves and the current-sheet breaks, the spectrum evolves towards a power law. At $t=181.48 \Omega_{e}^{-1}$ the energy cascades as $k_{\perp}^{-8/3}$ up to $k_{\perp}d_{e} \approx 4$. For $k_{\perp}d_{e} > 4$ the spectrum shortly steepens and then flattens at $k_{\perp}d_{e} \approx 10$. At later times, $t>181.48 \Omega_{e}^{-1}$, the spectrum flattens at $k_{\perp}d_{e} \approx 7$.} 
   
{Figure (\ref{fig:enterlabel13}b) shows $P^{B}_{1D_{\perp}} (k_{\perp}d_{e})^{8/3}$ for run $SC$. At $t=87.47 \Omega_{e}^{-1}$ there is a peak at $k_{\perp}d_{e} \approx 2$ which corresponds with the wavenumber of the driven fluctuations. The fluctuations slightly increase the magnetic energy at small-scales compared with run $SA$. Nevertheless, the evolution of the magnetic spectrum for run $SC$ is similar to run $SA$ and the spectrum flattens at $k_{\perp}d_{e} \approx 7$ at later times. The same is true for runs $SB-SH$ (not shown here).}  

{Figure (\ref{fig:enterlabel13}c) shows $P^{B}_{1D_{\perp}} (k_{\perp}d_{e})^{8/3}$ for run $SI$. For this run, the injected fluctuations increase the magnetic energy at $k_{\perp}d_{e} \approx 2$ and the spectrum forms a short plateau in the power spectrum with higher magnetic energy but with the same slope $k_{\perp}^{-8/3}$ at $t \geq 202.37 \Omega_{e}^{-1}$. At small-scales the spectrum flattens with time at $k_{\perp}d_{e} \approx 7$, similar to case $SA$.} 

{Figure (\ref{fig:enterlabel13}d) shows $P^{B}_{1D_{\perp}} (k_{\perp}d_{e})^{8/3}$ for run $SK$. For this run, at $t = 1.31 \Omega_{e}^{-1}$, the peak at $k_{\perp}d_{e} = 12.56$ shows the intense injection of energy. The spectrum is dominated by the injected energy and it saturates early. The slope for times $t = (110.97 - 335.53) \Omega_{e}^{-1}$ for $k_{\perp}d_{e} \geq 2$ corresponds to an uncompensated flat spectrum. The same is true for runs $SJ-SN$ (Not shown here).}    
%---------------------------------------------------------------------------------------------------------------------------------------------------------------

%---------------------------------------------------------------------------------------------------------------------------------------------------------------
Figure \ref{fig:parameterspace} summarizes the different runs in the driving frequency $\omega_{0}$ and wavenumber $k$ space of the driven fluctuations. The empty squares represent the runs in which the magnetic islands grow in time and the solid diamonds represent runs for which the magnetic island growing is suppressed. The data points are color-coded with $\langle |\delta B_{y} / \mathbf{B}| \rangle$. The vertical dashed-line marks the gyrofrequency $\Omega_{e}$ and the horizontal dashed-line marks gyroradious scale $k_{\rho_{e}}\sim \rho_{e}^{-1}$. 

For runs with small wavenumbers ($kd_{e} \leq 0.03$) the magnetic islands grow similar to run $SC$ in the range of frequencies $\omega_{0}/\omega_{pe} \sim (0.002 - 1)$ explored in this study for amplitudes $\langle |\delta B_{y}/\mathbf{B}| \rangle \leq 0.07$ (blue and green squares).

For runs with larger wavenumbers ($kd_{e} \sim 0.25$), only a large amplitude ($\langle |\delta B_{y}/\mathbf{B}| \rangle \approx 0.13 $) is able to disrupt the growth of the tearing mode (orange diamond). At shorter wavelengths ($k \sim d_{e}^{-1}$), fluctuations with amplitudes as small as ($\langle |\delta B_{y}/\mathbf{B}| \rangle \approx 0.07 $) are able to stop the growing of the magnetic islands (cyan diamonds). However, for fluctuations with $k \sim d_{e}^{-1}$ and smaller amplitudes ($\langle |\delta B_{y}/\mathbf{B}| \rangle \approx 0.01 $), the magnetic islands are able to grow (dark-blue square). This suggests the existence of an amplitude threshold $\langle |\delta B_{y,c}|/| \mathbf{B}|  \rangle$ that is spatial-scale dependent, {where $\alpha$ is a coefficient which can be scale dependent}.  

For instance, $\langle |\delta B_{y,c}|/| \mathbf{B}| \rangle = 0.13$ for $\lambda \sim \Delta$, whereas $\langle |\delta B_{y,c}|/| \mathbf{B}| \rangle = 0.069$ for $\lambda = d_{e}$ and $\langle |\delta B_{y,c}| /| \mathbf{B}| \rangle = 0.036$ for $\lambda = \rho_{e}$. For the range of frequencies that we explore in this study (Figure \ref{fig:parameterspace}), we do not observe temporal scale dependence for $\langle \delta B_{y,c} \rangle$.

\subsection{Effect of the driven fluctuations on the electron VDFs}

%-------------------------------------------------------------------
We now explore the effect of the driven fluctuations on the electron velocity distribution function (eVDF). 
We select four regions in the simulation domain. Figure \ref{fig:currentdensities}a) shows the out-of-plane current $J_{x}$ normalized to the max $J_{0}=max(|J_{x}|)$ for run $SB$ at $t_{1}=26.11 \Omega_{e}^{-1}$. Figure \ref{fig:currentdensities}b) shows $J_{x}/J_{0}$ for run $SB$ at $t_{2}=156.67 \Omega_{e}^{-1}$. At $t_{2}$ the current sheet is forming two magnetic islands and a dominant x-point. The black square (R1) highlights a region close to the dominant x-point, the red square (R2) highlights a region in the forming exhaust. The blue square (R3) and pink square (R4) highlight regions above the current sheet.

Figure \ref{fig:currentdensities}c) and \ref{fig:currentdensities}d) show $J_{x}/J_{0}$ for run $SJ$ at $t_{1}$ and $t_{2}$ respectively. At $t_{1}$, the current sheet is still present. However, at $t_{2}$ the current sheet has broken into pieces that remain of the same size for as long as the simulation lasts. For direct comparison we select the same regions as in Figure \ref{fig:currentdensities}a).

The top-row in Figure \ref{fig:VDFs} shows the electron VDFs for run $SB$ in the $v_{\parallel} - v_{\perp}$ plane with respect to the local magnetic field $\vec{B}_{l} = \langle \vec{B} \rangle$, where $\langle ... \rangle$ is the spatial average over the selected region highlighted by the squares in Figure \ref{fig:currentdensities}. The regions $R1$ and $R2$ are on the current sheet and the VDFs show a gyrotropic distribution. The VDFs at regions $R3$ and $R4$ are above the current sheet and show as well a gyrotropic distribution but with a lower core density which corresponds to the low background density. The electron VDF for run $SJ$ at $t_{1}$ in all four regions is similar to the results shown in top-row in Figure \ref{fig:VDFs}.

The middle-row in Figure \ref{fig:VDFs} shows isocontours of the electron VDF difference ($\Delta f = f_{t2} - f_{t1}$) for run $SB$, where $f_{t1}$ is the electron VDF at $t_{1}$ and $f_{t2}$ is the electron VDF at $t_{2}$. The red (blue) color represents $f_{t2} > f_{t1}$ ($f_{t2} < f_{t1}$). Within the current sheet, in regions $R1$ and $R2$, the electron VDF diffuses in an oblique angle in the anti-parallel direction as a result of the particle streams within the current sheet. Above the current sheet in regions $R3$ and $R4$, the electron VDF do not considerably changes from $t_{1}$ to $t_{2}$.      

The bottom-row in Figure \ref{fig:VDFs} shows $\Delta f = f_{t2} - f_{t1}$ for run $SJ$. Within the current sheet, in regions $R1$ and $R2$, the electron VDFs diffuse quasi-isotropically showing an increase in the thermal energy of electrons in both $v_{\parallel}$ and $v_{\perp}$. In contrast, above the current sheet in regions $R3$ and $R4$, the electron VDFs show a clear anisotropy in the perpendicular direction $T_{\perp}/T_{\parallel}>1$. This suggests that the effect of the driven fluctuations that are able to suppress the growing of magnetic islands is to increase the thermal energy within the current sheet, and to give preferential energy in the perpendicular direction to the electrons in the background population.   
%-------------------------------------------------------------------

%---------------------------------------------------------------------
\section{Discussion}
\label{sec:discussion}
%----------------------------------------------------------------
% Talk in terms of a thin enough current sheet. 
The run $SB$ in Figure \ref{fig:By_stack} shows the effect of the central pinch that speeds up the system to reach an early steady-state. Comparing runs $SE$ and $SF$ we conclude the reason for the delay is the higher guide field. 

For runs $SC$, $SD$ and $SE$ the overall dynamics is governed by the growing of the magnetic islands regardless of the broad band driven fluctuations. Moreover, our results suggest that only fluctuations with $\lambda \leq \Delta$ and with a scale-dependent critical magnetic amplitude $\delta B_{c}(kd_{e})$ are able to stop the growing of the magnetic islands and suppress the tearing mode.

{From Figure (\ref{fig:enterlabel13}), we observe that for growing magnetic-islands runs $SA-SH$ the driven magnetic fluctuations do not change the spectral evolution of the system and the energy transfer is dominated by the growing of the magnetic islands. The flattening of the spectrum is due to he saturation of magnetic energy and the absence of a strong dissipation mechanism at this stage. Although the kinetic and thermal energy of the particles increase with time (see figure (\ref{fig:enterlabel12})), there is still an accumulation of magnetic energy at small scales with time. In contrast, for all the non-growing magnetic-islands runs, ($SJ-SN$), the injected energy not only dominates the time evolution of the energy but also saturates the spectrum.}       

{Run $SI$ is an interesting case in which the magnetic-islands still grow, but the driven fluctuations modify the magnetic energy spectrum producing a plateau of higher magnetic energy but retaining the same slope. Since the range for this plateau is shorter than a decade in k-space, we let the interpretation of this effect for future studies. The non-dispersive nature of plasma waves below electron scales in pair-plasmas \citep{zocco2017slab, loureiro2018turbulence} poses an interesting challenge for the interpretation of possible mechanisms responsible for the dissipation of energy.}
%-------------------------------------------------------------------

%-------------------------------------------------------------------
In the absence of fluctuations, within the current sheet the slippage of magnetic field-lines induces an out of plane electric field $E_{x}\hat{x}$ that accelerates electrons along the local magnetic field introducing a perturbed out of plane current $\delta J_{x}$ \citep{drake1977kinetic}. This current sustains the magnetic field configuration associated with the magnetic islands. The out of plane current increases in time as shown by the solid-lines in Figure \ref{fig:currents_max}a). In contrast, the electric field fluctuations $\delta E_{x}(\omega,k)$ associated with the driven magnetic fluctuations are added to the induced electric field. These electric field fluctuations, at kinetic scales, remove the monotonous increase of the out of plane current as shown by the dashed-lines in Figure \ref{fig:currents_max}a). This is supported by the diffusion of the electron VDF into a hotter state as shown in the bottom row in Figure \ref{fig:VDFs}. The thermal motion overcomes the bulk flows needed to feed the expansion of the magnetic islands. This corresponds to the case in which the Doppler frequency $\omega_{D}$, associated with the thermal motion of electrons along the local magnetic field, is larger than the tearing growth rate  \citep{drake1977kinetic}.

The existence of a scale-dependent critical amplitude $\delta B_{y,c}$ can be associated with both wave-particle resonances and with non-resonant heating at small-scale wavelengths. At these scales, small-amplitude fluctuations change the local magnetic field associated with the particle gyration and therefore disrupt the particle orbits. For pair-plasma the properties of the cyclotron and mirror modes are expected to be similar to the electron-proton case \citep{gary2009fluctuations}; therefore some of the mechanisms that can explain this behaviour are the cyclotron resonance and the non-resonant stochastic heating \citep{chandran2010perpendicular}.

\section{Conclusions}
\label{sec:conclusions}
%------------------------------------------------------------

In this work we explore the effect of driving magnetic fluctuations at different spatial and temporal scales on a thin current sheet in a non-relativistic pair-plasma. We find that only magnetic fluctuations with $\lambda \leq \Delta$ and with an amplitude larger than a scale-dependent threshold amplitude $\delta B_{y,c}(\lambda)$ are able to suppress the growing of magnetic islands. Moreover, these fluctuations provide preferential perpendicular energy to the particles in the background population and increase the thermal energy of the particles within the current sheet. The anisotropy in the electron VDF of the background population introduced by the driven fluctuations might be able to trigger instabilities. {We let for future work the exploration of the scale dependence of $\delta B_{y,c}$ with $\lambda$. Moreover, the extension of our results to higher plasma beta is a natural continuation for this work.}

Although we drive fluctuations in an artificial manner, similar fluctuations at kinetic scales might arise from localized sources. For instance, at the exhaust and dipolarization fronts of preexisting reconnection events, lower hybrid waves, kinetic Alfvén waves (KAW) and whistler waves have been observed for ion-electron plasmas \citep{khotyaintsev2019collisionless} and inertial Alfvén waves might be present for pair-plasma. Moreover, the presence of a crescent feature on the agyrotropic electron VDF can trigger unstable upper hybrid waves that can affect the electron dynamics within the EDR \citep{graham2017instability,burch2019high}. These are high frequency waves that self-consistently formed by the reconnection dynamics.

We observe that the growing of plasmoids is a dominant process in the dynamics of a thin current sheet and it is insensitive to background turbulence at large scales ($\lambda > d_{e},\rho_{e}$) for a pair-plasma. Thus, our results cast some doubt on the role of turbulence as a trigger of multiple local reconnection events on a thin current sheet.

With regards to the role of these driven fluctuations on the reconnection mediated cascade \citep{boldyrev2017magnetohydrodynamic,loureiro2018turbulence}, these high spatial and temporal frequency fluctuations destroy the current sheets and break them into pieces with no formation of inverse cascade as there is no formation of magnetic islands. Thus, if fluctuations of this type are triggered by kinetic modes, the energy cascade might be modified. 

Our results might be relevant for magnetic confinement in laboratory plasmas as this mechanism suppress the growing of magnetic islands and of reconnection that can lead to saw-tooth crash associated with magnetic reconnection \citep{drake1991collisionless,yamada1994investigation,yamada1997study,bardoczi2017impact}. Previous experimental results also show the suppression of sawtooth crash by electron-cyclotron resonance heating \citep{hanada1991sawtooth}.

%------------------------------------------------------------
%\appendix
%---------------------------------------------------------------------

%\section{Appendix A}
%------------------------------------------------------------

%-------------------------------------------------------------------
\acknowledgments

%\section*{Funding}
%Details of all funding sources should be provided, including grant numbers if applicable. Please ensure to add all necessary funding information, as after publication this is no longer possible.

J.A.A.R.is supported by NASA grant 80NSSC21K2048 and NSF grant 2142430.

%---------------------------------------------------------------------

%---------------------------------------------------------------------
\typeout{}
\bibliography{Harris_current_with_driving_fluctuations}{}
\bibliographystyle{aasjournal}
%---------------------------------------------------------------------

\end{document}